\pdfoutput=1
\documentclass[10pt,a4paper]{article}
\usepackage{jheppub,bbold,graphicx,enumitem, hyperref, listings, multirow, subfig}
\usepackage[dvipsnames]{xcolor}
\usepackage[utf8]{inputenc}
\usepackage[normalem]{ulem}
\hypersetup{colorlinks=true, linkcolor=Maroon, citecolor=Maroon, filecolor=Maroon, urlcolor=Maroon}

\newcommand{\be}{\begin{equation}}
\newcommand{\ee}{\end{equation}}
\def\bsp#1\esp{\begin{split}#1\end{split}}
\newcommand{\ie}{\textit{i.e.}}

\newcommand{\amc}{{\sc MadGraph5}\_a{\sc MC@NLO}}
\newcommand{\fr}{{\sc Feyn\-Rules}}
\newcommand{\nloct}{{\sc NloCT}}
\newcommand{\mspin}{{\sc MadSpin}}
\newcommand{\mw}{{\sc MadWidth}}
\newcommand{\ma}{{\sc MadAnalysis~5}}
\newcommand{\py}{{\sc Pythia~8}}
\newcommand{\fj}{{\sc FastJet}}
\newcommand{\sfs}{{\sc SFS}}
\newcommand{\pyhf}{{\sc PyHF}}
\newcommand{\scr}[1]{\ensuremath{\mathcal{#1}}}

\def\sing{S_1}
\def\oct{S_8}

\newcommand*{\UCBL}{~\small \textit{Universit\'e Claude Bernard Lyon 1, CNRS/IN2P3,
Institut de Physique des 2 Infinis de Lyon, UMR 5822, F-69622, Villeurbanne, France}}
\newcommand*{\CPthree}{~\small \textit{Centre  for  Cosmology,  Particle  Physics  and  Phenomenology  (CP3), Université  catholique  de  Louvain,  1348  Louvain-la-Neuve,  Belgium}}
\newcommand*{\LPTHE}{~\small \textit{Laboratoire de Physique Th\'{e}orique et Hautes \'{E}nergies (LPTHE), UMR 7589,\\ Sorbonne Universit\'{e} \& CNRS, 4 place Jussieu, 75252 Paris Cedex 05, France}}
\newcommand*{\SYSU}{~\small  \textit{School of Physics, Sun Yat-Sen University, Guangzhou 510275, P. R. China}}

\title{Searching for top-philic heavy resonances in boosted four-top final states}

\author[a]{Luc Darmé,}
\affiliation[a]{\UCBL}
\emailAdd{l.darme@ip2i.in2p3.fr}
\author[b]{\! Benjamin Fuks,}
\emailAdd{fuks@lpthe.jussieu.fr}
\affiliation[b]{\LPTHE}
\author[c,d]{\! Hao-Lin Li,}
\affiliation[c]{\SYSU}
\emailAdd{lihlin68@mail.sysu.edu.cn}
\author[d]{\! Matteo Maltoni}
\emailAdd{matteo.maltoni@uclouvain.be}
\author[d]{\! and Julien~Touchèque}
\affiliation[d]{\CPthree}
\emailAdd{julien.toucheque@uclouvain.be}

\date{\today}

\abstract{
  New heavy resonances with sizeable couplings to top quarks can be probed through searches for beyond-the-Standard-Model effects in four-top production at the LHC. In this work, we present the first next-to-leading-order QCD predictions for the full on-shell and off-shell production of four-top events via new electroweak singlet states, along with dedicated analysis strategies based on the reconstruction and tagging of all final-state top quarks. We develop a detector-level simulation incorporating recent advances in top-tagging and boosted object reconstruction. Moreover, we demonstrate that searches at LHC Run~3 and high-luminosity phase in the zero-lepton, one-lepton and same-sign di-lepton channels can improve the sensitivity to the new physics cross sections by up to two orders of magnitude. In particular, colour-octet resonances with masses up to 2--2.5~TeV and colour-singlet states with masses up to 1--1.5~TeV are within reach for coupling values in the 0.1--1 range.
}

\begin{document}\preprint{IRMP-CP3-25-24}
\maketitle
\flushbottom

\section{Introduction}\label{sec:intro}

The ability to reconstruct the nature of new coloured particles from the detailed observation of jet substructure has become a cornerstone of many analyses at the LHC. In recent years, the deployment of machine-learning-based techniques that exploit the full set of jet constituents and their properties has led to significant advances in jet tagging performance, these developments representing one of the most exciting innovations of the latest experimental runs at the LHC~\cite{Kasieczka:2017nvn, Larkoski:2017jix, Guest:2018yhq}. By dramatically reducing mistagging rates by factors compared to traditional methods~\cite{Gerbush:2007fe, ATLAS:2015ddu, CMS:2017wyc, ATLAS:2018wis, ATLAS:2022qby, Cagnotta:2022hbi, ATLAS:2024rua}, these algorithms open the door to novel search strategies, especially in regimes where large Standard Model (SM) backgrounds have previously limited sensitivity. 
In light of these new approaches, final states with high object multiplicities, such as those originating from the production of multiple top quarks possibly induced by physics beyond the SM, are particularly promising targets. It has hence been found that in such scenarios, the ability to resolve and identify individual substructures within jets is crucial to improving signal efficiency and enabling robust discrimination from background processes~\cite{Englert:2016aei, Aguilar-Saavedra:2019ptp, Belyaev:2021zgq, Araz:2023axv, Chowdhury:2023jof, Darme:2024epi, Sahu:2024fzi}. 

Four-top final states represent one of the highest-multiplicity SM processes accessible at the LHC, with each top quark decaying to a $W$ boson and a $b$ quark. Owing to their complexity and rarity, four-top events have the potential to offer a powerful probe of both SM and beyond-the-SM (BSM) physics. Dedicated analyses from both ATLAS and CMS~\cite{CMS:2017ocm, ATLAS:2018alq, ATLAS:2018kxv, CMS:2019jsc, CMS:2019rvj, ATLAS:2020hpj, ATLAS:2021kqb, ATLAS:2023ajo, ATLAS:2023taw, CMS:2023ftu,CMS:2023zdh, ATLAS:2024jja} have hence developed increasingly sophisticated strategies, typically based on reconstructed objects such as $b$-tagged jets and isolated leptons.  These studies, leveraging data from the LHC Run~2 dataset, have provided valuable insights, notably enabling complementary constraints on the SM top Yukawa coupling through comparisons with $t\bar{t}H$ production. Until recently~\cite{ATLAS:2024jja}, most four-top searches have concentrated on the kinematic regime expected from SM predictions~\cite{Bevilacqua:2012em, Alwall:2014hca, Maltoni:2015ena, Frederix:2017wme, Jezo:2021smh, vanBeekveld:2022hty, Dimitrakopoulos:2024qib, Dimitrakopoulos:2024yjm, vanBeekveld:2025ghw}, characterised by relatively soft and isolated top decay products. However, this regime poses challenges for full event reconstruction due to the limited Lorentz boost of the individual tops, yielding a high combinatorial background associated with the large amount of well-isolated decay products of the four-top system. As a result, much of the available phase space, particularly that involving energetic and boosted top quarks as predicted in many BSM scenarios, remains under-explored.

In this context, a wide variety of new physics models predict final states with multiple top quarks, with four-top production emerging as a particularly compelling signature. In particular, such top-rich final states arise naturally when new heavy particles strongly coupled to the top quark are pair-produced via QCD interactions, and subsequently decay to top-antitop pairs~\cite{Battaglia:2010xq, Dev:2014yca, Greiner:2014qna, Alvarez:2016nrz, Kim:2016plm, Fox:2018ldq, Alvarez:2019uxp, Blasi:2023hvb}. Composite Higgs models constitute a prominent class of such setups, offering a solution to the hierarchy problem yielding composite resonances such as vector-like top partners or coloured pseudo-Nambu-Goldstone bosons with enhanced couplings to the top quark~\cite{Lillie:2007hd, Pomarol:2008bh, Zhou:2012dz, Cacciapaglia:2015eqa, Belyaev:2016ftv, Liu:2019bua, Cacciapaglia:2020vyf, Cornell:2020usb, Cacciapaglia:2024wdn}. Other well-motivated frameworks giving rise to four-top final states include minimal flavour violation models~\cite{Gerbush:2007fe, Hayreter:2017wra}, extended supersymmetries~\cite{Salam:1974xa, Fayet:1974pd, Fayet:1975yi, AlvarezGaume:1996mv, Fox:2002bu, Plehn:2008ae, Choi:2008ub, GoncalvesNetto:2012nt, Fuks:2012im, Calvet:2012rk, Benakli:2014cia, Beck:2015cga, Kotlarski:2016zhv, Darme:2018dvz, Carpenter:2020hyz, Carpenter:2020evo} and constructions with an extended Higgs sector~\cite{Branco:2011iw, Arcadi:2019lka, Cheng:2017tbn, Cheng:2018mkc, Coloretti:2023yyq, Anisha:2023xmh}. Additionally, recent phenomenological studies have also exploited four top-quark production in an effective field theoretical framework~\cite{Agram:2013koa, Khatibi:2014via, Durieux:2014xla, Guo:2016kea, Shen:2018mlj, DHondt:2018cww, Hartland:2019bjb, Degrande:2020evl, Liu:2020bem, Banelli:2020iau, Darme:2021gtt}. 

In this work, we focus on the direct production of heavy top-philic states decaying on-shell to top-antitop pairs and leading to a distinctive four-top signature. To capture the essential phenomenology while retaining model independence, we employ the same simplified model framework as the one developed in our earlier studies~\cite{Fuks:2012im, Calvet:2012rk, Beck:2015cga, Darme:2018dvz, Darme:2021gtt, Darme:2024epi}. These works have demonstrated that four-top final states provide a robust and complementary probe of top-philic sectors, both in the resonant regime and in an effective construction when the mediators are too heavy to be produced. As highlighted in our previous analysis~\cite{Darme:2024epi}, reconstructing and tagging all four boosted top quarks in the final state dramatically suppresses the Standard Model background and opens the door to novel and more sensitive search strategies, notably in the zero-lepton, one-lepton and same-sign di-lepton channels. In particular, we have found that using modern top-tagging techniques enables a potential improvement of up to an order of magnitude over existing limits, especially in the regime of high-mass resonances. This enhanced sensitivity results from two dominant effects: a significantly increased signal efficiency in the boosted-top regime compared to traditional $b$-jet-based selections targetting a resolved four-top signal, and a strong reduction of the irreducible QCD background allowing the exploration of the more challenging fully hadronic final states.
This strategy parallels recent approaches developed in the context of four-bottom final states for di-Higgs production~\cite{ATLAS:2022hwc, CMS:2024ymd}, where the full reconstruction of boosted $b$-jets systems has yielded substantial gains in sensitivity. It also connects naturally with recent theory-driven efforts to apply machine-learning techniques to improve four-top searches with a jet substructure analysis~\cite{Choudhury:2024mox, Kvita:2024ooa, Flacke:2025xwl}. 

We further present in this article the first complete next-to-leading order (NLO) projections for the signal, offering a more realistic estimate of the experimental sensitivity to BSM-induced four-top production. The signal predictions are obtained using custom UFO~\cite{Degrande:2011ua, Darme:2023jdn} models for simplified top-philic scenarios, developed with the {\sc MoGRe} framework~\cite{Frixione:2019fxg} and recent extensions of \fr~\cite{Christensen:2009jx, Alloul:2013bka, Degrande:2014vpa, Frixione:2019fxg}. To our knowledge, this constitutes the first theoretical NLO estimate of new physics processes leading to four-top final states including both resonant and non-resonant contributions. We find that the NLO cross sections can exceed the leading-order (LO) predictions by up to 75\% in certain benchmark scenarios, significantly enhancing the projected reach of future LHC analyses. This highlights the importance of consistent NLO modelling when forecasting the sensitivity of multi-top searches. For the background, we place particular emphasis on the precise simulation of the $t\bar{t} + \text{jets}$ QCD background, which remains the dominant contribution. To accurately model it, we implement a full matching and merging scheme with multi-leg LO matrix elements, ensuring a reliable and consistent description of both the hard scattering and the parton shower processes. This is found essential for faithfully accounting for jets that may fake boosted top quarks. Consequently, appropriate background rejection procedure could be put in place.

In Section~\ref{sec:theory}, we outline the main theoretical foundation of our work and define the simplified models used throughout our analysis, and Section~\ref{sec:boost} details the simulation framework and the reconstruction techniques employed to target a boosted four-top final state. In Section~\ref{sec:ana}, we present our background modelling and describe the analysis strategies used to probe both colour-singlet and colour-octet mediator signals. Our main results and projected LHC sensitivities are reported in Section~\ref{sec:results}. Finally, we conclude in Section~\ref{sec:conclu}, whereas additional technical details on the NLO simulation setup and analysis strategy are provided in Appendices~\ref{sec:app} and \ref{appendixDiffDist}.

\section{Theoretical and numerical framework}\label{sec:theory}

In this section, we present the theoretical and numerical framework used throughout our analysis. We focus on two classes of simplified models that extend the SM with top-philic scalar particles: a colour-singlet scalar $\sing$ state of mass $M_{\sing}$ and a colour-octet scalar $\oct$ state of mass $M_{\oct}$. These simplified extensions provide a minimal and general framework for capturing the key features of a broad class of UV completions such as those mentioned in Section~\ref{sec:intro}, and the corresponding interactions can be described by a compact Lagrangian formulation which we introduce in Section~\ref{sec:lags} along with a selection of representative benchmark points. We then exploit the fact that the simplified model approach offers a flexible path for phenomenological interpretation and numerical simulation in Section~\ref{sec:evtgen}. To this aim, we implement these models in the UFO format~\cite{Degrande:2011ua, Darme:2023jdn} using the \fr~\cite{Christensen:2009jx, Alloul:2013bka}, {\sc MoGRe}~\cite{Frixione:2019fxg} and \nloct~\cite{Degrande:2014vpa} packages. This setup then allows us to perform signal simulations at NLO in QCD, the corresponding cross sections and $K$-factors computed using \amc~\cite{Alwall:2014hca} being discussed in detail.

\subsection{Simplified models for top-philic new physics}\label{sec:lags}

The new physics contributions to the Lagrangian of the two simplified models considered here and featuring top-philic scalar resonances include three terms: gauge-invariant kinetic and mass terms for the new resonance, as well as a Yukawa-like coupling to a top-antitop pair. This yields the singlet and octet Lagrangians $\mathcal{L}_{\sing}$ and $\mathcal{L}_{\oct}$ given by
\begin{equation}\label{eq:lags}\bsp
    \scr{L}_{\sing} =&\ \scr{L}_{\mathrm{SM}} + \frac{1}{2} \partial_\mu \sing \partial^\mu \sing- \frac{1}{2} M_{\sing}^2 \sing^2 + y_{\sing} \sing  \, \bar{t} t \ ,\\
  \scr{L}_{\oct} =&\  \scr{L}_{\mathrm{SM}} + \frac{1}{2} D_\mu \oct^A D^\mu \oct^A - \frac{1}{2} M_{\oct}^2 \oct^A \oct^A + y_{\oct}  \oct^A  \, \bar{t}\, T^A\, t  \ .
\esp\end{equation}
where the additional scalar fields are assumed to be real and $A$ indicates (summed) colour-adjoint indices. Here, $\scr{L}_{\mathrm{SM}}$ denotes the SM Lagrangian, $y_{\sing}$ and $y_{\oct}$ are the new Yukawa couplings, while $M_{\sing}$ and $M_{\oct}$ represent the masses of the singlet and octet states respectively.

Each of these simplified models can be naturally connected to UV-complete new physics constructions. In particular, colour-octet states, regardless of their spin or CP quantum numbers, frequently arise in UV models addressing the hierarchy problem of the SM. For example, minimal supersymmetric scenarios feature colour-octet fermions known as gluinos which have been extensively searched for at the LHC and that are now constrained to be heavier than approximately 2~TeV, depending on the specific model. In extended supersymmetric frameworks, gluinos belong to supermultiplets that also include scalar or pseudo-scalar octet fields (generally organised into complex scalar fields) commonly referred to as sgluons. These scalar resonances have attracted considerable attention as they can alleviate the tension that too heavy gluinos pose on the Higgs sector~\cite{Salam:1974xa, Fayet:1974pd, Fayet:1975yi, AlvarezGaume:1996mv, Fox:2002bu, Plehn:2008ae, Choi:2008ub, GoncalvesNetto:2012nt, Fuks:2012im, Calvet:2012rk, Benakli:2014cia, Beck:2015cga, Kotlarski:2016zhv, Darme:2018dvz, Carpenter:2020hyz, Carpenter:2020evo}. If the pseudo-scalar octet is the lightest coloured BSM particle, then it is stable at tree level and couples to the SM quarks at one loop with a strength proportional to the quark mass, thus becoming top-philic. Composite models offer another motivation for top-philic scalar resonances. Analogous to QCD, such models predict meson-like composite states arising from a new strongly-coupled gauge sector, and scenarios typically exhibit a rich spectrum of bound states including pseudo-scalar colour-charged mesons that emerge as pseudo Nambu-Goldstone bosons below the new confinement scale (much like the pions in QCD)~\cite{Lillie:2007hd, Pomarol:2008bh, Zhou:2012dz, Belyaev:2016ftv, Cacciapaglia:2015eqa, Liu:2019bua, Cacciapaglia:2020vyf, Cornell:2020usb, Cacciapaglia:2024wdn}. While a wide variety of couplings are possible, sizeable interactions with the top quark can often arise~\cite{Belyaev:2016ftv}. In addition, colour-singlet scalars also commonly appear in such extensions, as well as in a broad class of other new physics models.

More generally, the BSM top-quark couplings introduced in eq.~\eqref{eq:lags} are expected to be generated, in a UV-complete model, only after electroweak symmetry breaking (EWSB). Since the Higgs vacuum expectation value is the sole source of EWSB in the SM, it is reasonable to expect that any scalar singlet not involved in EWSB would inherit a coupling structure to quarks proportional to their SM Yukawa couplings, thereby favouring a top-philic scenario. This mechanism is in fact common in dark sector constructions~\cite{Arcadi:2019lka}. Similarly, constructions with an extended Higgs sector such as the Two-Higgs-Doublet Model, stringent flavour constraints often enforce alignment in the Yukawa sector~\cite{Branco:2011iw}, again leading to enhanced couplings to the top quark.

Finally, we note that as the scalar or pseudo-scalar nature of the top-philic resonance does not affect the analysis proposed in this study, we focus on scalar top-philic states in the remainder of this work. Our results would nevertheless remain valid if the scalar interactions in eq.~\eqref{eq:lags} were replaced with pseudo-scalar ones. Likewise, although vector particles would lead to different projected bounds on their Lagrangian parameters, the analysis strategy presented here would remain applicable. We leave a dedicated study of alternative spin and parity assignments, as well as of the colour-sextet case which shares many features with the octet scenario at the analysis level, for future work.

\begin{figure}
  \centering
  \includegraphics[width=0.3\textwidth]{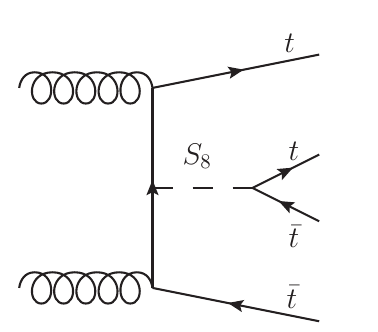}
  \includegraphics[width=0.3\textwidth]{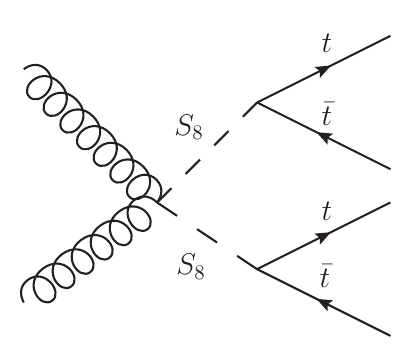}
  \caption{Representative Feynman diagrams contributing to four-top production in a simplified model with an additional scalar octet field $\oct$. When the scalar particles are on-shell, the associated production of the scalar octet with a top-antitop pair (left) is proportional to the square of the Yukawa coupling from the second Lagrangian of eq.~\eqref{eq:lags}, while the QCD-driven pair production of two scalar octets (right) depends only on the octet mass $M_{\oct}$ after assuming that $\mathrm{Br}(\oct \to t\bar{t}) = 1$. \label{fig:FeynDiagram}}
\end{figure}

From a bottom-up perspective, the main phenomenological difference between the two simplified models in eq.~\eqref{eq:lags} lies in the possibility of producing pairs of colour-octet states via QCD interactions, as illustrated by the representative Feynman diagrams on the right of Figure~\ref{fig:FeynDiagram}. Apart from this, in both cases the new physics particle is expected to predominantly decay into top quarks, leading to a signal comprising four top quarks. The projections that will be made below can however be straightforwardly rescaled to account for a reduced branching ratio, as may occur in next-to-minimal or UV-complete models featuring multiple decay modes. The parameter space of both simplified models is thus two-dimensional and defined by the scalar mass $M_X$ and the Yukawa-like coupling $y_X$ with $X = \oct$ or $\sing$. For the colour-octet case, the dominant production mechanism (associated or pair production, respectively illustrated in the left and right of Figure~\ref{fig:FeynDiagram}) depends sensitively on the values of these parameters. As the mass $M_{\oct}$ increases, on-shell production becomes suppressed and the relative importance of pair production diminishes compared to the associated production. A similar effect arises when increasing the Yukawa coupling $y_{\oct}$, as associated production scales with $y_{\oct}^2$ whereas pair production remains dominated by QCD.

To explore these features, we define four benchmark points (BPs) that probe different regions of the parameter space and highlight contrasting topologies in four-top production. For the colour-octet model, we consider:
\begin{itemize}[topsep=2pt,itemsep=0pt,parsep=3pt,partopsep=2pt]
    \item \textbf{BP1:} $M_{\oct} = 2$~TeV, $y_{\oct} = 1$ (yielding $\Gamma_{\oct} = 38$~GeV);
    \item \textbf{BP2:} $M_{\oct} = 3$~TeV, $y_{\oct} = 3$ (yielding $\Gamma_{\oct} = 527$~GeV).
\end{itemize}
BP1 corresponds to an intermediate case where neither production mechanism dominates, allowing us to study a mixed production regime. In contrast, BP2 implies an increased associated production contribution while pushing the particle width into a regime ($\Gamma_{\oct} / M_{\oct} \sim 0.18$) where finite-width effects and off-shell kinematics become non-negligible~\cite{Denner:1999gp, Denner:2005fg}. For the colour-singlet scenario in which only the Yukawa-induced associated production is present, we define two benchmarks focused on lower and intermediate masses with Yukawa couplings of $\mathcal{O}(1)$, reflecting the limited reach of the LHC for colour-neutral resonances:
\begin{itemize}[topsep=2pt,itemsep=0pt,parsep=3pt,partopsep=2pt]
    \item \textbf{BP3:} $M_{\sing} = 1.5$~TeV, $y_{\sing} = 1$ (with $\Gamma_{\sing} = 165$~GeV);
    \item \textbf{BP4:} $M_{\sing} = 2$~TeV, $y_{\sing} = 1.5$ (with $\Gamma_{\sing} = 513$~GeV).
\end{itemize}
The adopted mass range $M_X \sim 1-3$~TeV is particularly relevant for LHC searches, as most coloured BSM particles are now excluded below $\sim 2$~TeV in simplified scenarios. Colour-octet scalars are a notable exception with current constraints reaching only up to about $1.3$~TeV~\cite{Darme:2018dvz, ATLAS:2024jja}. As we will show, this limit can be significantly improved by efficiently tagging the boosted top quarks arising from heavy scalar decays.

\subsection{Event generation tool chain}\label{sec:evtgen}
As illustrated by the representative Feynman diagrams in Figure~\ref{fig:FeynDiagram}, the set of BSM four-top production processes considered in this work is
\begin{equation}\label{eq:processes}
\begin{cases}
    & pp \to t\bar{t} X \to t\bar{t}t\bar{t} \quad \text{with } X = \sing \text{ or } \oct, \\
    & pp \to \oct \oct \to t\bar{t}t\bar{t}.
\end{cases}
\end{equation}
In the following calculations, we neglect all electroweak-induced amplitudes involving, for instance, Higgs or $W/Z$ boson exchange as they are expected to be subdominant relative to the new physics or QCD contributions. Moreover, due to the high multiplicity of coloured final state particles, NLO QCD corrections are anticipated to be significant and are therefore included in the modelling of the BSM signal for all channels. This section details the simulation pipeline used to generate the corresponding BSM four-top signal event samples at NLO accuracy in QCD.

As a first step, we implement the simplified models defined by the Lagrangians of eq.~\eqref{eq:lags} in the UFO format~\cite{Degrande:2011ua, Darme:2023jdn}. The model implementation is carried out with \fr~\cite{Christensen:2009jx, Alloul:2013bka}, starting from a reduced version of the SM Lagrangian containing only terms involving QCD-charged fields, the top Yukawa interaction, and additionally assuming a unit CKM matrix and five active quark flavours. The Lagrangian is next extended with the top-philic scalar interactions of eq.~\eqref{eq:lags}, and we generate two UFO models, one for each scalar representation ($\sing$ and $\oct$). For consistency with the modified version of \amc~\cite{Alwall:2014hca} described in Appendices~\ref{appendixrenoFULL} and \ref{appendixrenoQCD}, the new particles are assigned PDG code 9000001, although this choice can be adjusted if needed provided that the required changes in \amc\ are modified accordingly.  

The fields are renormalised in the on-shell scheme and the couplings in the $\overline{\mathrm{MS}}$ scheme. For the strong coupling, we use a scheme with five active quark flavours such that the top-quark and the new coloured state loops in the gluon self-energy are subtracted at zero momentum transfer, while the contributions of the light quarks and gluons are treated in the $\overline{\mathrm{MS}}$ scheme. In this way, the running of $\alpha_s$ is generated solely by the gluon and the light-quark flavours. This renormalisation procedure is handled using {\sc MoGRe} (introduced in Appendix~B of Ref.~\cite{Frixione:2019fxg}), to renormalise the QCD sector of the SM as well as the new BSM interactions. By construction, electroweak bosons and the SM Higgs are unaffected by this renormalisation procedure as their interactions are absent from the SM sector of the Lagrangian. This choice enables a fully consistent NLO QCD computation of BSM four-top production, and is justified by previous findings~\cite{Darme:2021gtt} showing that interferences of new physics and electroweak amplitudes become phenomenologically relevant only for very heavy resonances and/or in the non-perturbative regime of the model, where an effective field theory description would be more appropriate than a resonance-based one. We emphasise that the generation of a suitable UFO library requires the \lstinline{dev-bsm} version of \fr,\footnote{See \url{https://github.com/FeynRules/FeynRules/tree/feynrules-dev-bsm}.} as the \lstinline{current} version does not support our renormalisation procedure.

To generate the one-loop counterterms and $R_2$ rational terms required for NLO calculations in four dimensions, we make use of \nloct~\cite{Degrande:2014vpa} and {\sc FeynArts}~\cite{Hahn:2000kx}. Since the BSM scalars always decay into top pairs, we assume throughout that the scalar mass satisfies $M_X > 2 m_t$ with $X=\sing$ or $\oct$. Moreover, in order to consistently describe unstable particles with potentially large widths in our simulation chain, we adopt the complex mass scheme~\cite{Denner:1999gp,Denner:2005fg} and produce UFO models accordingly, ensuring that gauge invariance is preserved in all undertaken calculations.

Hard-scattering events are then generated at NLO in QCD using \amc, with virtual corrections handled via the \textsc{MadLoop} routines~\cite{Hirschi:2011pa}. However, the latter discard by default all loop diagrams involving the scalar singlet as it is not coloured. This is problematic since we renormalised both the QCD and BSM sectors of the model so that the set of generated loop diagrams should include contributions involving the $\sing$ resonance. We consequently modify the default \amc\ behaviour following the procedure of Refs.~\cite{Borschensky:2020hot, Borschensky:2021hbo}, also detailed in Appendix~\ref{appendixrenoFULL} for the models considered in this study. This leads to a consistent cancellation of both IR and UV poles across the virtual, counterterms and real-emission contributions. In contrast, for the scalar octet case no modification to \amc\ is required as the particle is colour-charged and automatically included in the generated loop diagrams. For readers interested in using standard UFO models built with the default QCD renormalisation provided by \nloct\ and \fr, we outline in Appendix~\ref{appendixrenoQCD} why this approach, which we do not use in this work, generally fails for BSM processes involving off-shell particles coupling to quarks. We also describe the necessary modifications to achieve consistent IR and UV divergence cancellation in this case. We have explicitly verified that both approaches (the full QCD+BSM renormalisation and the minimal QCD-only one) lead to identical NLO cross sections across our benchmark points. This reflects the fact that the additional diagrams included in the extended QCD+BSM renormalisation approach correspond to subleading corrections.

In addition, at large scalar octet masses and couplings, the numerical reduction of loop integrals becomes increasingly delicate. In this regime, we observe occasional failures of the pole cancellation checks at specific phase-space points. We traced these instabilities to the default usage of the {\sc Collier} library~\cite{Denner:2016kdg} in {\sc MadLoop}. To overcome this, we switch to alternative reduction tools such as {\sc Ninja}~\cite{Peraro:2014cba} and {\sc CutTools}~\cite{Ossola:2007ax}.

In all simulations relevant for our study, we use the NNPDF2.3NLO set~\cite{Ball:2012cx, Ball:2013hta} of parton distribution functions (PDF) and estimate theoretical uncertainties via variations in the renormalisation and factorisation scales and PDF replicas. The central scale is fixed to half the total hadronic activity in the event $H_T/2$, and we apply a minimum transverse momentum cut of $p_T\!>\!10$~GeV on all parton-level jets. Once parton-level events are generated, we model the inclusive decays of the top quarks using \mspin~\cite{Artoisenet:2012st} and \mw~\cite{Alwall:2014bza}, and the resulting decayed events are then matched to parton showers and hadronisation as implemented in \py~\cite{Bierlich:2022pfr} using the default parameters.\footnote{\py\ is also used for the handling of inclusive tau-lepton decays.} While this final step has a very limited impact on the BSM signal due to the high transverse momentum of the tops produced in the BSM particle decays, it turns out to be essential for accurately modelling the dominant $t\bar{t}+$jets background that requires the extra modifications discussed in Section~\ref{sec:bkd}. In total, we generate fully showered and hadronised NLO signal event samples comprising 300,000 events for each benchmark point considered. In addition, we prepare a sparse grid of 18 intermediate points with 10k events each to enable interpolation and limit setting following the procedure described in Section~\ref{sec:results}. All model files and Monte Carlo configuration cards are publicly available on Zenodo~\cite{darme_2025_15783920}.


\subsection{Signal NLO cross sections and K-factors}\label{sec:Kfact}
Running large event samples at NLO accuracy demands significant computational resources. We therefore restrict full NLO computations to the benchmark points introduced in Section~\ref{sec:lags} and validate that the corresponding NLO signal distributions remain sufficiently close to their LO counterparts. This enables us to rely on LO simulations to derive constraints while correcting the total rate using $K$-factors. The relevant kinematic distributions, discussed in Section~\ref{sec:ana}, confirm that this approximation is justified: LO shapes provide a reliable estimate of the signal features while the NLO effects after the selection are largely captured by a global normalisation factor. In this approach, the approximate number of selected BSM events at NLO accuracy ($N_{\rm K-NLO}$) is given by
\begin{align}\label{eq:multscheme}
  N_{\rm K-NLO} \equiv \mathcal{L} \, \varepsilon_{\rm LO} \, \left( \frac{ \sigma_{\rm NLO} }{\sigma_{\rm LO} } \right) \, \sigma_{\rm LO} \equiv \mathcal{L} \, \varepsilon_{\rm LO} \, K \, \sigma_{\rm LO} \ ,
\end{align}
where $\mathcal{L}$ denotes the integrated luminosity, $\varepsilon_{\rm LO}$ is the selection efficiency evaluated using LO simulations and $K$ is defined as the ratio of the NLO and LO cross sections $\sigma_{\rm NLO}$ and $\sigma_{\rm LO}$ when they are computed using renormalisation and factorisation scales set to the mass of the BSM resonance. The strength of this method lies in its computational efficiency: both $\varepsilon_{\rm LO}$ and $K$ can be determined across a sparse grid in the mass/coupling parameter space at a fraction of the cost required for full NLO event generation, and then fitted linearly. The LO cross section $\sigma_{\rm LO}$ is finally evaluated on a finer grid since it can be obtained with reduced computational resources. Moreover, this approach is motivated by the observation that both the efficiencies and the $K$-factors exhibit slow variation in terms of the model's parameters compared to the total production cross section.

\begin{figure}
    \centering
    \includegraphics[width=\linewidth]{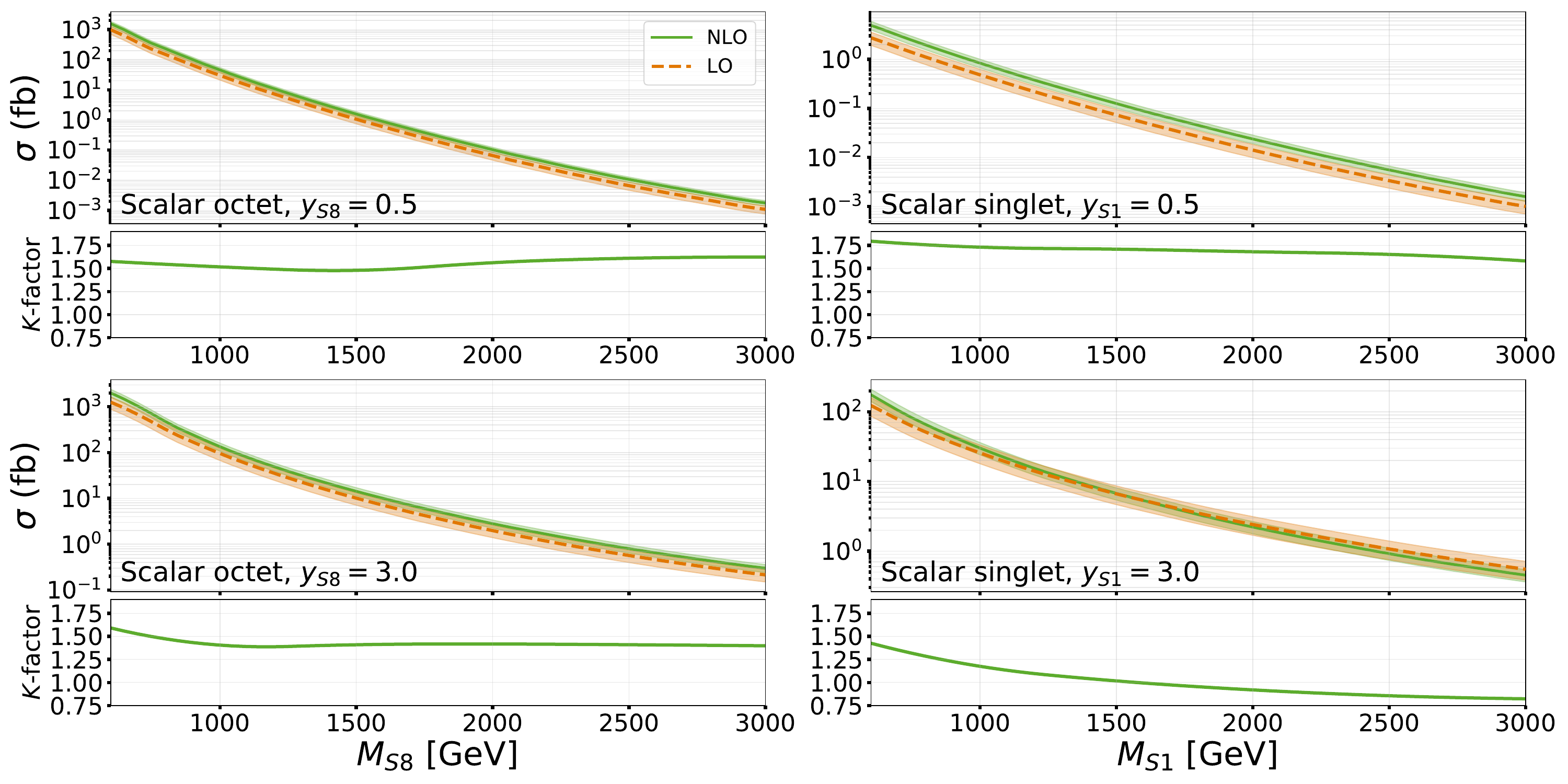}
    \caption{Comparison of LO and NLO cross sections as functions of the BSM resonance mass. Results are shown for the scalar octet ({\it left}) and singlet ({\it right}) cases, assuming couplings to top quarks of 0.5 ({\it top}) and 3 ({\it bottom}) . The lower panels of the figures display the associated $K$-factors defined as the ratio of the NLO cross section to the LO prediction computed with NLO PDFs.}
    \label{fig:CS}
\end{figure}

Figure~\ref{fig:CS} shows the four-top production cross sections induced by the inclusion in the field content of the theory of a scalar octet (left) and a scalar singlet (right), for two specific values of the corresponding Yukawa coupling to the top quark that we choose equal to 0.5 (top row of the figure; weak coupling) and 3 (bottom row of the figure; strong coupling). The shaded bands indicate theoretical uncertainties estimated using the standard seven-point scale variation procedure: the renormalisation and factorisation scales are independently varied by factors of two around the central scale, excluding the extreme combinations where one scale is multiplied by 0.5 and the other by 2. This conventional definition of scale uncertainties is further adopted throughout the rest of this work. In agreement with the results of~\cite{Darme:2021gtt}, we find that QCD-induced scalar octet pair production, thus independent of the top-quark coupling to the new resonance as illustrated with the representative diagram shown in Figure~\ref{fig:FeynDiagram}, dominates the total cross section up to resonance masses of about 2~TeV. Beyond this point, phase-space suppression becomes significant and the less-suppressed contribution from associated production overtakes, before eventually dominating at higher masses. For the scalar singlet where only associated production contributes, the total cross section therefore falls more slowly with increasing mass compared to the octet case. Finally, across the entire region of the parameter space experimentally accessible at the LHC, the $K$-factors, that we show in the lower panels of the different figures, typically lie in the range $1.4-1.8$. These moderately large values show that NLO corrections can play a critical role in improving the reliability of the total rate prediction. Altogether, after accounting for the modest effect on the shapes of the signal differential distributions (see Section~\ref{sec:ana}), our hybrid approach using LO shapes with NLO-corrected normalisation is found to offer an optimal balance between accuracy and computational cost for the parameter scans required in our study.

\section{Characterisation of the boosted four-top system} \label{sec:boost}

\subsection{Simulation of the detector response and object definitions}\label{sec:objects}
We simulate the response of a typical LHC detector using the \sfs\ framework~\cite{Araz:2020lnp, Araz:2021akd} implemented within \ma~\cite{Conte:2012fm, Conte:2014zja, Conte:2018vmg}. Since the reconstruction of top quarks plays a central role in this study, we recalibrated the default ATLAS detector parametrisation to improve agreement with reference experimental studies~\cite{ATLAS:2018rvc, ATLAS:2019qmc, ATLAS:2020lks}, focusing in particular on the invariant mass reconstruction of top-quark candidates. Electrons and muons are reconstructed following the medium working-point performance described in Refs.~\cite{ATLAS:2019qmc, ATLAS:2020auj}, respectively. Additionally, two jet collections are defined, both using a reconstruction based on the anti-$k_T$ algorithm~\cite{Cacciari:2008gp} as implemented in \fj~\cite{Cacciari:2011ma}, but with two different radius parameters $R=0.4$ (AK4 jets) and $R=1.0$ (AK10 jets). For AK4 jets, $b$-tagging is applied probabilistically using the $p_T$- and $\eta$-dependent efficiencies of Ref.~\cite{ATLAS:2016gsw}. To avoid double-counting between the AK4 jet and lepton collections, we implement a series of overlap removal procedures following the prescription of Ref.~\cite{ATLAS:2020hpj}. AK4 jets are first removed if they are within $\Delta R < 0.2$ of a lepton, although in the case of a muon the jet must also have three or fewer associated tracks. Next, electrons and muons are respectively removed if they lie within $\Delta R < 0.4$ or $0.04 + 10\,\mathrm{GeV}/p_T$ of any of the remaining AK4 jets, while finally electrons within $\Delta R < 0.1$ of a muon are also discarded.

For boosted top-quark identification, we apply different tagging strategies depending on the context. For the validation of our reconstruction procedure, we use the jet-mass and N-subjettiness classifier from Ref.~\cite{ATLAS:2015ddu}, while for our main four-top analysis, we rely on more advanced constituent-based top-tagging algorithms applicable to AK10 jets~\cite{ATLAS:2022qby}. In this last case, we specifically adopt the performance of the \lstinline{HlDNN} and \lstinline{ParticleNet} classifiers for jets satisfying $p_T > 350$~GeV, $|\eta| < 2.0$ and invariant mass $M_j > 40$~GeV. AK10 jets matched to a partonic top within $\Delta R < 0.75$ are top-tagged with an efficiency of 80\%, whereas jets not consistent with a top quark are mis-tagged at an average rate of either 10\% (conservative, \lstinline{HlDNN}-like) or 5\% (optimistic, \lstinline{ParticleNet}-like), the exact value depending on the jet $p_T$.

In our validation procedure, we have obtained excellent agreement with the expectations of an ATLAS search for resonant $t\bar{t}$ production in the semi-leptonic channel~\cite{ATLAS:2018rvc}, recovering the reconstructed resonance mass distribution, the signal selection efficiencies and the exclusion limits within 20\% for both the resolved and boosted signal regions of the ATLAS analysis. We have also recovered the fact that for $t\bar{t}$ resonance masses lying between 1 and 2~TeV, the production rate at the LHC is high enough that systematic uncertainties dominate. Subsequently, potential improvements in the higher-luminosity LHC runs are not foreseen. Moreover, as observed in Refs.~\cite{Carena:2016npr, Djouadi:2019cbm} and more recently in a CMS public note~\cite{CMS:2025rnx}, interference effects between the SM and BSM amplitudes for gluon-initiated $t\bar{t}$ production become significant above 1~TeV. These effects can create a dip rather than a peak in the top-antitop invariant mass distribution, complicating the statistical interpretation of the search. These two limitations provide an additional motivation for studying four-top final states as a probe of top-philic new physics in the high-mass regime. In particular, in this channel, the impact of interference is indeed negligible within the parameter space of interest, allowing a more robust interpretation of potential deviations from the SM prediction.

We now turn to detailing in the following subsection the analysis selection criteria imposed on the reconstructed objects introduced here will be detailed.

\subsection{Reconstruction of a boosted four-top system} \label{sec:topreco}
\begin{figure}
    \centering
    \includegraphics[width=\linewidth]{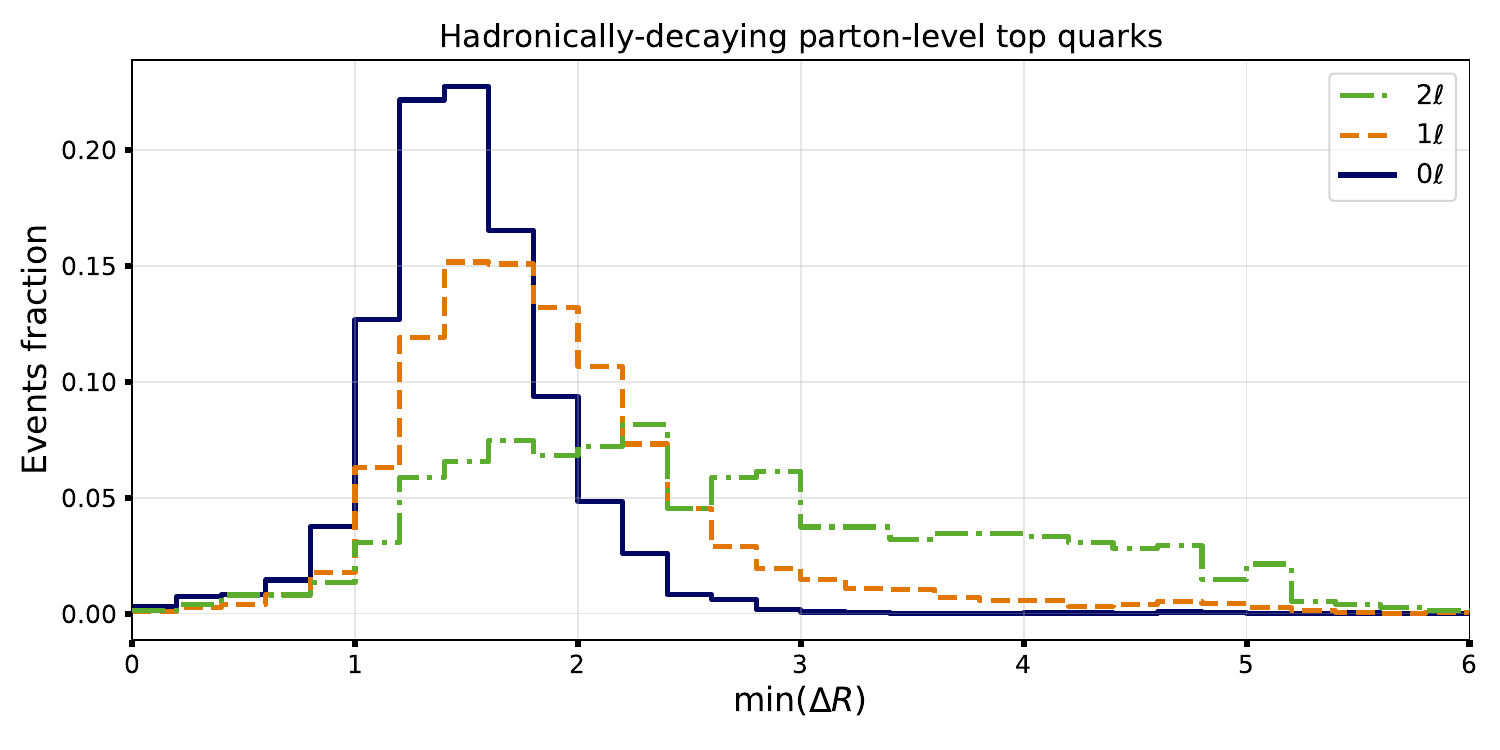}
    \caption{Minimum $\Delta R$ separation between (parton-level) top quarks, computed for events that pass at least one signal region selection after reconstruction. Events are categorised according to the number of leptons produced in the top quark decays.}
    \label{fig:minDR_tHadr}
\end{figure}

Since our main interest is the exploration of BSM resonances with masses above the TeV scale, the resulting top quarks are typically boosted enough to be accurately reconstructed and tagged using standard top-tagging algorithms such as those described in Section~\ref{sec:objects}. We have verified that for BSM resonances in the TeV range, thus viable relatively to current exclusion limits, our signal samples feature four top quarks that tend to decay in isolation from each other. As a result, the reconstruction procedure validated on BSM-induced top-antitop events remains reliable even in the case of a higher-multiplicity $t\bar{t}t\bar{t}$ final state. In particular, since AK10 jets clustered by \fj\ serve as proxies for hadronically-decaying top quarks, we have checked that the typical angular distance $\Delta R$ between them remains larger than 1, allowing thus for the independent reconstruction of each top jet. As shown in Figure~\ref{fig:minDR_tHadr}, fewer than 5\% of the selected events contain two hadronic top quarks separated by $\Delta R < 1$ at parton level.

Based on this observation, we implement a set of selection criteria starting from the classification of events according to the number of leptons and AK10 jets that they contain. Leptons must satisfy $p_T > 20$~GeV and $|\eta| < 2.5$ to be considered as potential decay products of leptonically-decaying top quarks. In addition, all reconstructed AK10 jets are ordered by decreasing $p_T$ while giving priority to those that are top-tagged. However, any AK10 jet overlapping with an isolated lepton within $\Delta R < 1$ is discarded. $b$-jets included in the reconstruction are taken to be AK4 jets with $p_T > 25$~GeV and $|\eta| < 2.4$ that pass our $b$-tagging requirements. We refer to these $b$-jets as `isolated' when they do not overlap with any of the selected AK10 jets within $\Delta R < 1$ so that they will be allowed to serve as ingredients for the reconstruction of leptonically-decaying top quarks. Non-$b$-tagged AK4 jets are retained only if they have $p_T > 40$~GeV and $|\eta| < 2.4$. The missing transverse momentum $p_T^\text{miss}$ is calculated as the negative vector sum of the transverse momenta of all leptons and jets (both $b$-tagged and non-$b$-tagged) that pass the above cuts, thereby minimising contamination from neutrinos produced in the parton shower~\cite{Darme:2024epi}. At this stage, light jets are thus only used in the $p_T^\text{miss}$ calculation.

The subsequent reconstruction procedure depends on the number of isolated leptons present. If no such lepton is found, the event is considered fully hadronic. For the analysis targeting the colour-octet case, at least four AK10 jets are required without imposing top-tagging requirements. For the colour singlet case, we instead require at least two AK10 jets along with a minimum of two isolated $b$-jets in order to suppress background.

When exactly one isolated lepton is present, the event is interpreted as containing a single leptonically-decaying top quark and three hadronically-decaying ones. In the colour octet case, at least three AK10 jets are required to represent the three hadronic tops, along with at least one isolated $b$-tagged AK4 jet. In the singlet case, we instead again require at least two AK10 jets and at least two isolated $b$-jets. In both scenarios, the isolated $b$-jet closest in $\Delta R$ to the lepton is selected for reconstructing the leptonic top. In addition, we require $p_T^\text{miss} > 25$~GeV in consistency with the presence of a neutrino. The longitudinal component of this neutrino is then estimated by assuming that it is produced together with the lepton in the decay of an on-shell $W$ boson. Up to two real solutions may be obtained from the kinematic fit resulting from this assumption, and we select the one that yields a reconstructed top quark mass closest to the expected value when the $W$ boson and $b$-jet momenta are combined. If no real solution is available, we retain the real part of the complex solutions to proceed.

In events with two isolated leptons, we require them to have the same electric charge to suppress the dominant $t\bar{t}+$jets background. We further demand two AK10 jets to account for the two hadronically-decaying tops and two isolated $b$-jets to reconstruct the leptonically-decaying ones. If more than two such $b$-jets are available, the ones closest in $\Delta R$ to the leptons are used. Furthermore, we require $p_T^\text{miss} > 50$~GeV and estimate the transverse momenta of the two neutrinos under the assumption that the event originates from the decay of heavy parent particles, which tends to favour configurations featuring a low transverse mass when combining the leptons and the neutrinos~\cite{Lester:1999tx}. Letting $(\ell_1, \nu_1)$ and $(\ell_2, \nu_2)$ denote the lepton-neutrino pairs, we define the transverse mass of each $W$ boson as
\begin{equation}
    M_T^{(i)} = \sqrt{2 |\vec{p}_T^{\,\ell_i}| |\vec{p}_T^{\,\nu_i}| \, (1-\cos\Delta \phi_{\ell\nu})},
\end{equation}
where $i = \{1, 2\}$ and $\Delta \phi_{\ell\nu}$ is the azimuthal angle between the lepton and the neutrino. The unknown neutrino transverse components are estimated by using the so-called stransverse mass $M_{T2}$~\cite{Lester:1999tx, Cheng:2008hk} that is defined through the minimisation condition
\begin{equation}
    M_{T2} = \min_{\vec{p}_T^{\,\nu_1} + \vec{p}_T^{\,\nu_2} = \vec{p}_T^\text{miss}} \max\Big(M_T^{(1)}, M_T^{(2)}\Big).
\end{equation}
The longitudinal components of the two neutrinos are then inferred from kinematic fits assuming an on-shell $W$ boson decay. Among all solutions obtained for each leptonic top quark, we select the one that leads to a reconstructed top mass closest to the physical value, using the $b$-jet closest in $\Delta R$ to the associated lepton. As before, if no real root exists, we use the real part of the complex solution.

\begin{table}[t]
  \setlength\tabcolsep{10pt}\renewcommand{\arraystretch}{1.2}
  \resizebox{0.98\textwidth}{!}{\begin{tabular}{cccccc} \hline
    \multicolumn{6}{l}{\rule{0pt}{1.25em}\textbf{Basic kinematic requirements}} \\[0.75em]
      & Electrons & Muons & AK4 Jets& AK10 Jets& $b$-jets \\
      $p_T$ (GeV) & $>20$ &  $>20$ &  $>20$ &$>350$ &  $>25$ \\
      $|\eta|$ & $<2.47$ &  $<2.5$ &  $<2.5$ & $<2.0$ & $<2.4$ \\[0.75em]
   \hline
    \multicolumn{6}{l}{\textbf{\rule{0pt}{1.25em}Object definitions}} \\[0.75em]
    Non $b$-tagged AK4 jets & \multicolumn{5}{|l}{$p_T > 40 $ GeV, $|\eta| \le 2.4$} \\
    Isolated leptons & \multicolumn{5}{|l}{$p_T \ge 20$ GeV, $|\eta| \le 2.5$}\\
    Isolated AK4 jets & \multicolumn{5}{|l}{ $\Delta R > 1$ from any AK10 jet} \\
    Missing energy & \multicolumn{5}{|l}{$p_T^{\textrm{miss}} > 25$ GeV (one isolated lepton) or 50 GeV (two isolated leptons)}\\[0.75em]
  \hline
    \multicolumn{6}{l}{\textbf{\rule{0pt}{1.25em}Boosted four-top selection for colour-octet resonances}}\\[0.75em]
    No isolated lepton & \multicolumn{5}{|l}{$\ge4$ AK10 jets} \\
    One isolated lepton & \multicolumn{5}{|l}{$\ge3$ AK10 jets, $\ge1$  isolated $b$-tagged AK4 jet} \\
    Two isolated leptons & \multicolumn{5}{|l}{$\ge2$ AK10 jets, $\ge2$ isolated $b$-tagged AK4 jet} \\[0.75em]
    
    \multicolumn{6}{l}{\textbf{\rule{0pt}{1.25em}Boosted four-top selection for colour-singlet resonances}}\\[0.75em]
    $\leq$ Two leptons & \multicolumn{5}{|l}{$\ge2$ AK10 jets, $\ge2$ isolated $b$-tagged AK4 jet} \\[0.75em]
  \hline
  \end{tabular}}
    \caption{Summary of the preselection cuts. See main text for details. \label{tab:preselection} }
\end{table}

At the end of this procedure that we summarise in Table~\ref{tab:preselection}, each event is associated with a set of reconstructed top quark candidates classified according to their decay modes (leptonic or hadronic) and top-tagging status. These are thus ready to be used in further analysis steps that we will describe in Section~\ref{sec:bkd}.

\section{Backgrounds and analysis strategy}\label{sec:ana}
As outlined in the previous sections, we develop two distinct analyses targeting the production of four top quarks via intermediate colour-octet and colour-singlet resonances. In the colour-octet scenario, pair production of resonances can dominate the cross section in certain regions of the parameter space, requiring the reconstruction of four objects that serve as proxies for the four top quarks produced in the resonance decays. In the colour-singlet case, only a single resonance is produced, allowing the selection to solely focus on two boosted top quark candidates, while the remaining two top quarks are expected to arise from QCD interactions and thus to feature different properties. Finally, in both scenarios, we also implement a same-sign dilepton selection strategy, which provides a complementary handle on the signal albeit with reduced efficiency due to the low branching ratio of this final state.

\subsection{Signal region definition and pairing strategy in the colour-octet model} \label{sec:octetana}
\begin{table}\renewcommand{\arraystretch}{1.3}\setlength\tabcolsep{14pt}
    \centering
    \begin{tabular}{c|cccc}
        \multicolumn{5}{c}{Colour-octet analysis} \\ \hline
        Signal Region & \# $\ell$& \# $b_\ell$ & \# AK10 & \# top-tag. \\ \hline
        \multirow{2}{*}{SR1} & 0 &  - & $\ge4$ & $\ge3$ \\
                             & 1 &  $\ge1$ & $\ge3$ & $\ge2$ \\ \hline
        \multirow{2}{*}{SR2} & 0 & - & $\ge4$ & $\ge4$ \\
                             & 1 & $\ge1$ & $\ge3$ & $\ge3$ \\ \hline
        SSL & 2 (same-sign) &  $\ge2$ & - & - \\
    \end{tabular}
    \caption{Summary of the selection criteria defining each signal region in the colour-octet analysis. The table lists the required number of isolated leptons $\ell$, the number of isolated AK4 $b$-jets associated with leptonically-decaying top quarks $b_\ell$, as well as the number of AK10 jets and top-tagged AK10 jets. We remind that AK10 jets are imposed not to overlap with any isolated lepton within $\Delta R = 1$, while isolated $b$-jets are similarly defined as not overlapping with any AK10 jet within the same angular distance.}
    \label{tab:octet_requirements}
\end{table}

In the search strategy designed for the colour-octet simplified model, we aim to reconstruct two on-shell BSM particles of equal mass by pairing four identified top quark candidates. When an event contains four such objects, it is assigned to one or more overlapping signal regions based on the number of top-tagged AK10 jets.
\begin{itemize}[topsep=2pt,itemsep=0pt,parsep=3pt,partopsep=2pt]
    \item The \textbf{SR1} region includes fully hadronic events with at least three top-tagged AK10 jets, as well as single-lepton events featuring at least two top-tagged AK10 jets. This region thus gathers events with at least three reconstructed and tagged (hadronic or leptonic) top quarks, without any top-tagging requirements on the fourth AK10 jet.
    \item The \textbf{SR2} region is defined by stricter conditions so that four top quarks are reconstructed and tagged. Fully hadronic events must thus contain at least four top-tagged AK10 jets, while one-leptonic events must feature at least three top-tagged AK10 jets.
    \item The \textbf{SSL} region encompasses events with two leptons of the same electric charge, benefiting hence from reduced background contamination so that no additional top-tagging of any AK10 jet is required.
\end{itemize}
A summary of these requirements is provided in Table~\ref{tab:octet_requirements}, which also highlights that SR1 and SR2 are not mutually exclusive: any events satisfying SR2 conditions automatically populate the SR1 region too.

Each selected event thus contains four reconstructed objects. Some are explicitly top-like, like for instance a leptonically-decaying top or a top-tagged AK10 jet, while others may be less clearly identified, like a non-tagged AK10 jet. These four objects are enforced to be paired into two groups to estimate the mass of the resonances that might have produced them. While experimental searches often use machine-learning techniques to optimise this pairing (like a boosted decision tree in a recent ATLAS study~\cite{ATLAS:2022hwc}), our goal here is to provide a simple and transparent illustration of the sensitivity of four-top final states to BSM top-philic resonances. Therefore, we adopt a minimalistic invariant-mass matching approach, similar in spirit to the distance metric employed by CMS in~\cite{CMS:2024ymd}. This may seem surprising as in principle, both the pair and associated production mechanisms compete for the colour-octet model, with relative rates depending on the underlying model parameters. However, while these two topologies can be distinguished at parton level, the reconstruction process tends to smear their kinematic features, making this distinction practically ineffective at reconstructed level (see Appendix~\ref{appendixDiffDist} for a further discussion on this point). For this reason, we choose to consistently pair the four top candidates by minimising the absolute difference between the two invariant masses of the pair, an approach equivalent to assuming pure pair production. This strategy, that is analogous to the one implemented in standard di-Higgs searches where each Higgs boson decays into a $b\bar{b}$ pair, will be justified \textit{a posteriori} by the competitive bounds that we will derive on the total BSM-induced four-top production cross section. 

\begin{figure}
    \centering
    \includegraphics[width=\linewidth]{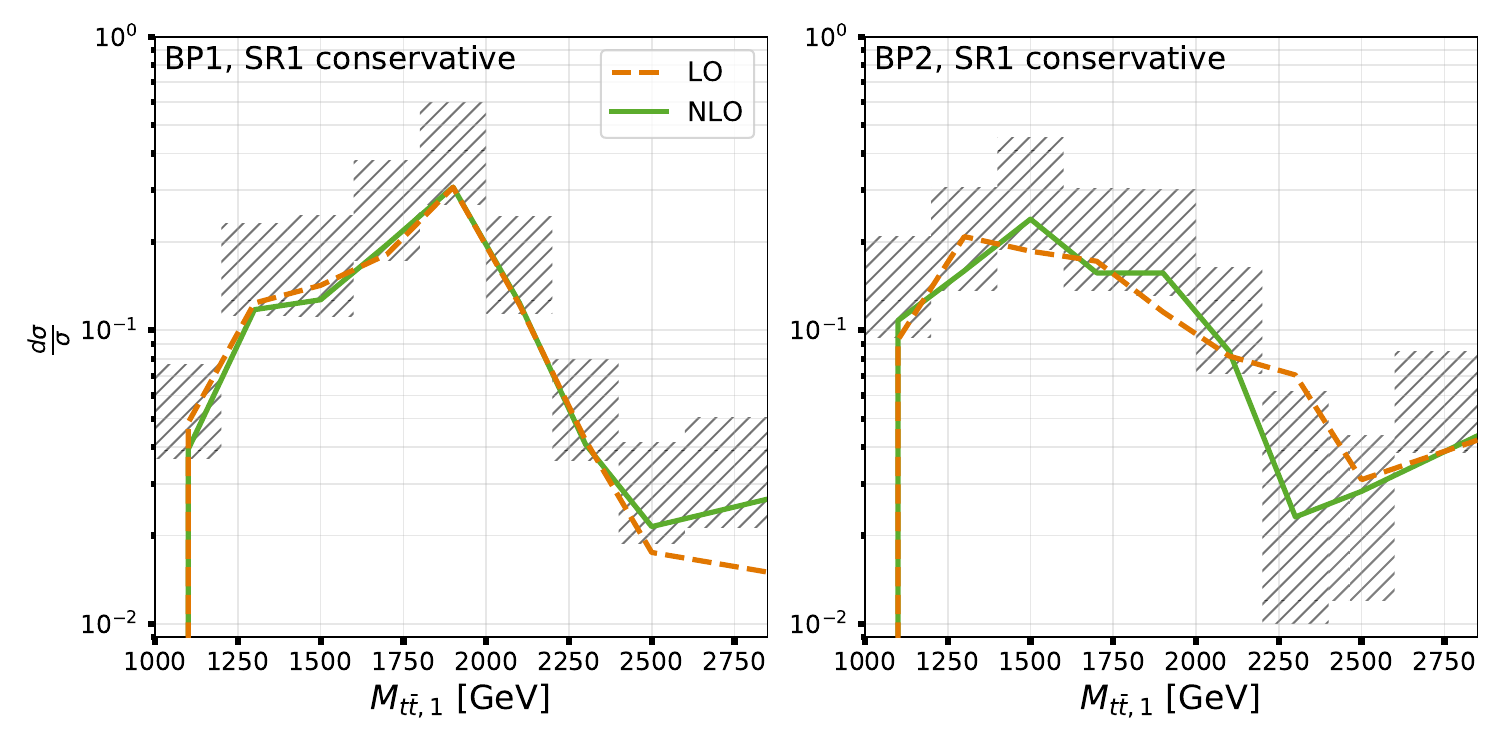}
    \caption{LO (dashed orange) and NLO (solid green) distribution of the largest reconstructed resonance mass for the colour-octet benchmark points BP1 (left) and BP2 (right), for the SR1 signal region with conservative top-tagging performance. The predictions are normalised to 1, the hatched bands represent the NLO scale variation and statistical uncertainties and the last bin includes the overflow. \label{fig:loVsNlo_octet}}
\end{figure}

In addition, no information about the top quark electric charge is used during the pairing since this information is not experimentally accessible. The sole exception is in the SSL region, where the presence of two same-sign leptons ensures they cannot originate from the same resonance, yielding thus a natural constraint to impose during pairing. Due to momentum smearing, the two reconstructed resonance masses typically differ. We denote them by $M_{tt,1}$ and $M_{tt,2}$, ordered such that $M_{tt,1} > M_{tt,2}$. We then use $M_{tt,2}$ for further background suppression by requiring $M_{tt,2} > 1$~TeV. Only events passing this cut contribute to the $M_{tt,1}$ distribution that we will further exploit, after binning it in 200~GeV intervals ranging from 1 to 2.6~TeV with the final bin including any overflow.

Since our reconstruction and selection efficiencies are derived from LO simulations (see Section~\ref{sec:Kfact}), it is crucial to verify that LO and NLO predictions remain consistent. Figure~\ref{fig:loVsNlo_octet} shows the $M_{tt,1}$ distributions at LO and NLO for the two colour-octet benchmark scenarios BP1 and BP2 defined in Section~\ref{sec:lags}. The two shapes are compatible within the NLO scale uncertainties shown as hatched bands. While minor differences can be seen in the tails of the distributions, they are not significant and solely reflect the limited statistics of our NLO Monte Carlo samples.

\subsection{Signal region definition and resonance reconstruction in the colour-singlet model}\label{sec:singletana}
\begin{table}[t] \renewcommand{\arraystretch}{1.3}\setlength\tabcolsep{14pt}
    \centering
    \begin{tabular}{c|cccc}
        \multicolumn{5}{c}{Colour-singlet analysis} \\ \hline
        Signal Region & \# $\ell$ &  \# $b_\ell$ & \# AK10 & \# top-tag. \\ \hline
        \multirow{2}{*}{SR1} & 0 &  $\ge2$ & $\ge2$ & $\ge1$ \\
                             & 1 & $\ge2$ & $\ge2$ & $\ge1$ \\ \hline
        \multirow{2}{*}{SR2} & 0 & $\ge2$ & $\ge2$ & $\ge2$ \\
                             & 1 &  $\ge2$ & $\ge2$ & $\ge2$ \\ \hline
        SSL & 2 (same-sign) &  $\ge2$ & - & - \\
    \end{tabular}
    \caption{Summary of the selection criteria defining each signal region in the colour-singlet analysis. The table lists the required number of isolated leptons $\ell$, the number of isolated AK4 $b$-jets associated with leptonically-decaying top quarks $b_\ell$, as well as the number of AK10 jets and top-tagged AK10 jets. We remind that AK10 jets are imposed not to overlap with any isolated lepton within $\Delta R = 1$, while isolated $b$-jets are similarly defined as not overlapping with any AK10 jet within the same angular distance. \label{tab:singlet_requirements}}
\end{table}

The search strategy targeting the colour-singlet model builds on the fact that only one pair of top quarks is expected to originate from the decay of a heavy resonance. As a consequence, only two top quark candidates are required to reconstruct the BSM resonance mass, and we correspondingly ask for at least two AK10 jets in the event preselection regardless of the lepton multiplicity. If more than two AK10 jets are found, we focus on the two with the highest transverse momentum, giving precedence to those that are top-tagged, and these two AK10 jets then serve as proxies for hadronically-decaying top quarks. 

As detailed in Section~\ref{sec:topreco}, to further suppress the background we additionally require each event to contain at least two isolated $b$-tagged jets (defined as jets not overlapping with any AK10 jets within a distance $\Delta R = 1$). These $b$-jets may then be used to reconstruct leptonically-decaying top quarks, depending on the lepton content of the event. More precisely, in fully hadronic events (zero leptons), the two selected AK10 jets are directly used for the reconstruction of the BSM resonance. In the single-lepton case, the isolated $b$-jet closest in $\Delta R$ to the lepton is combined with it to reconstruct a leptonically-decaying top quark following the strategy outline above and relying on a kinematic fit of the event. Finally, in the dilepton case, each lepton is paired with its nearest $b$-tagged jet to reconstruct two such objects, the four-momenta of the two neutrinos being reconstructed from the $M_{T2}$-based strategy discussed previously.

Signal regions are next defined based on the number of top-tagged AK10 jets observed.
\begin{itemize}[topsep=2pt,itemsep=0pt,parsep=3pt,partopsep=2pt]
    \item The \textbf{SR1} region includes all fully-hadronic and single-lepton events with at least one top-tagged AK10 jet. These events thus contain at least one reconstructed and tagged object accompanied by either an additional non-tagged AK10 jet or a leptonically-decaying top quark.
    \item The \textbf{SR2} region contains events with at least two top-tagged AK10 jets, independently of the number of leptons or other non-tagged AK10 jets.
    \item The \textbf{SSL} region includes all same-sign dilepton events, regardless of the number of tagged AK10 jets.
\end{itemize}
A summary of the requirements for each signal region is given in Table~\ref{tab:singlet_requirements}.

Once the preselection and signal region assignment are completed, we proceed to reconstruct the resonance mass. We pair the two top quark candidates with the highest transverse momentum (with precedence being given to the top-tagged AK10 jets), assuming that they originate from the BSM resonance decay in an associated production topology. Their invariant mass is finally computed and stored in a histogram. For the SR1 and SR2 regions, we use bins of 200~GeV from 400 to 4000~GeV with the last bin containing the overflow, while for the SSL region the range is limited to 1600~GeV. Since four top-like objects are not always available in these events, we do not attempt to compute a second invariant mass or apply any additional background rejection based on extra top candidates.

\begin{figure}
    \centering
    \includegraphics[width=\linewidth]{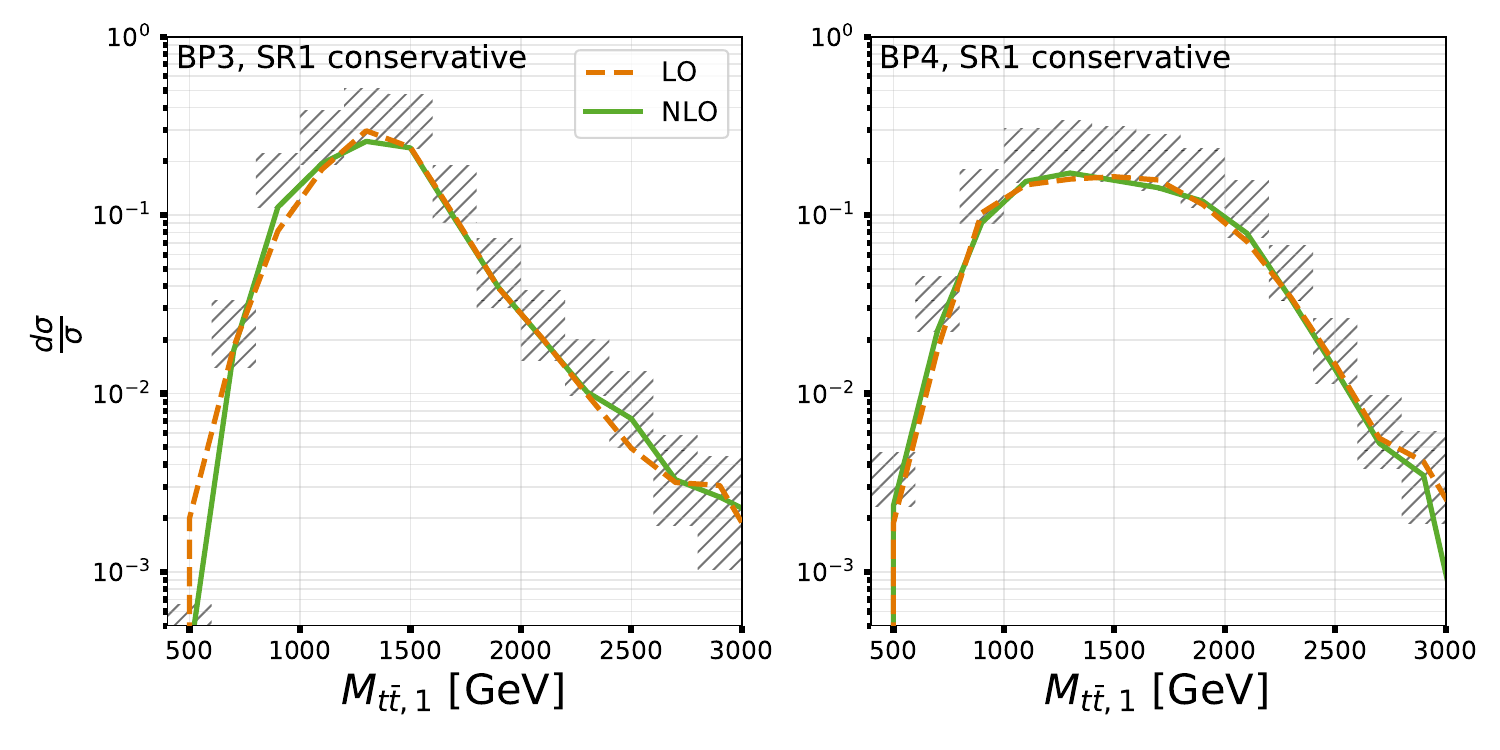}
    \caption{LO (dashed orange) and NLO (solid green) distributions of the reconstructed resonance mass for the colour-singlet benchmark points BP3 (left) and BP4 (right), for the SR1 signal region with conservative top-tagging performance. The predicted differential cross sections $\mathrm{d}\sigma$ are normalised to 1, the hatched bands represent the NLO scale variation and statistical uncertainties and the last bin includes the overflow. \label{fig:loVsNlo_singlet}}
\end{figure}

As discussed in Section~\ref{sec:Kfact}, reconstruction and selection efficiencies will be obtained from LO simulations. It is therefore important to validate that LO and NLO differential distributions are in reasonable agreement. To this aim, the reconstructed resonance invariant mass distributions at LO and NLO for the benchmark points BP3 and BP4 are presented in Figure~\ref{fig:loVsNlo_singlet}. The shapes of the LO and NLO distributions  are consistent within the scale variation uncertainties of the NLO prediction, showing that the LO simulation strategy introduced in Section~\ref{sec:lags} can be safely used. As for the octet case, the deviations observed in the high-mass tails are not significant, and are attributable to limited statistics in the NLO Monte Carlo samples.

\subsection{Background description} \label{sec:bkd}

As described in the previous sections, our signal selection strategy is primarily based on the reconstruction and the identification of boosted top quarks. This contrasts with existing experimental four-top searches, including the most recent analyses by ATLAS~\cite{ATLAS:2024jja} and CMS~\cite{CMS:2023ftu}, which do not rely on this feature. As a result, the relevant background composition differs significantly and requires a dedicated reassessment as compared to \cite{CMS:2023ftu, ATLAS:2024jja}.

\begin{table}
\centering\renewcommand{\arraystretch}{1.3}\setlength\tabcolsep{14pt}
\begin{tabular}{c|ccc|ccc}
Process & $\sigma$(LO) & Scale & PDF & $\sigma$(NLO) & Scale & PDF \\ 
\hline
$t \bar{t} j j $ & 354 &  $^{+ 62\%}_{- 35\%}$ & $\pm~ 5.8\%$ &
   352 & $^{+ 3.7\%}_{- 13\%}$ & $\pm~ 2.6\%$ \\   
$t \bar{t} W $ & 0.376 & $^{+23\%}_{-17\%}$ & $\pm~3.9\%$ & 
   0.565 & $^{+ 8.3\%}_{- 8.3\%}$ & $\pm~ 1.8\%$ \\
$t \bar{t} W j $ & 0.329 & $^{+39\%}_{-26\%}$ & $\pm~2.1\%$ & 
   0.452  & $^{+ 8.1\%}_{- 12\%}$ & $\pm~ 1.2\%$ \\
$t \bar{t} Z $ & 0.563 &  $^{+ 31\%}_{- 22\%}$ & $\pm~4.8\%$ &
   0.756 & $^{+ 9.2\%}_{- 11\%}$ & $\pm~ 2.1\%$ \\
$t \bar{t} Z j $ & 0.639 & $^{+ 47\%}_{- 30\%}$ & $\pm~ 6.5\%$ & 
    0.672 & $^{+ 2.6\%}_{- 9\%}$ & $\pm~ 2.5\%$ \\
$t \bar{t} t \bar{t}$ & 0.00612 & $^{+ 65\%}_{- 37\%}$ & $\pm~ 13\%$ & 
   0.00920 & $^{+ 28\%}_{- 24\%}$ & $\pm~ 6.0\%$ \\
   
$t \bar{t} t + t\bar{t}\bar{t}$ & 0.00155  & $^{+ 22\%}_{-17\%}$ & $\pm~ 13\%$ &
   0.00201 &  $^{+20 \%}_{-19 \%}$ &  $\pm~7.5 \%$ \\
\end{tabular}
\caption{LO and NLO cross sections (in pb) for the dominant SM background contributions relevant to our analysis, computed at LO and NLO in QCD for a centre-of-mass energy of 13~TeV. The predictions are obtained using the NNPDF2.3NLO parton density set~\cite{Ball:2012cx}, and include theory uncertainties from scale and PDF variations. The renormalisation and factorisation scales are centrally set to half the total hadronic transverse energy in the event ($H_T/2$), and a minimal transverse momentum of $p_T > 20$~GeV is required for each parton-level jet.\label{tab:Xsec}}
\end{table}

The dominant background in our study arises from the production of a top-antitop pair in association with additional jets and/or electroweak bosons (that are decayed inclusively in our simulation chain). In particular, the process $pp \to t \bar{t} jj$ where the extra jets can mimic boosted top quarks constitutes its main contribution. Subleading background components include the $pp \to t \bar{t} V$ and $pp \to t \bar{t} Vj$ processes with $V = W, Z$ that we treat independently due to our analysis requirements. The latter indeed favour contributions where additional jets are highly energetic jets and could be mistagged as top quarks. As such jets are better modelled at the matrix-element level and not by parton showering, this allows us to consider two non-overlapping background samples for the $pp \to t \bar{t} V$ and $pp \to t \bar{t} Vj$ processes. SM four-top ($pp \to t \bar{t} t \bar{t}$) and three-top ($pp \to t \bar{t} t$, $t \bar{t} \bar{t}$) production have cross sections at least two orders of magnitude smaller. Despite their reduced impact, we include these processes for completeness. Conversely, we have verified that multijet and $t \bar{t} VV$ backgrounds become negligible after applying our selection and top-tagging procedure. Cross sections for the most relevant background processes, both at LO and NLO in QCD, are collected in Table~\ref{tab:Xsec} for a centre-of-mass energy $\sqrt{s} = 13$ TeV.

Background simulations are achieved with the toolchain introduced earlier but with a differing configuration for the colour-octet and colour-singlet analyses. For the octet case, we require at least three final-state parton-level objects (\ie\ prior to decay) with $p_T > 300$ GeV to enhance the chance of reconstructing four boosted top proxies, tagged or not. In the singlet case where only two top candidates are needed, we instead impose this condition on just two parton-level objects. All background samples are generated at LO due to computational constraints, as the strong background rejection stemming from our analysis selections makes the generation of sufficient statistics at NLO particularly costly and not so needed. However, we have explicitly verified that the NLO cross sections remain compatible with their LO counterparts (as listed in Table~\ref{tab:Xsec}), at least when the hard $p_T$ cuts are replaced by a minimal cut of 20~GeV.

\begin{table}[t]
    \centering\renewcommand{\arraystretch}{1.5}\setlength{\tabcolsep}{6pt}
    \resizebox{\textwidth}{!}{
    \begin{tabular}{c|ccccccc}
       \multicolumn{8}{c}{Colour-octet selection [fb]} \\
         & $t \bar t +$jets & $t \bar t W$ & $t \bar t Wj$ & $t \bar t Z$ & $t \bar t Zj$ & $t \bar t t \bar t$ & $t \bar t t + t \bar t \bar t$ \\ \hline
       SR1 & 5.8$\cdot 10^{-2}{}^{\ +22\%}_{\ -17\%}$ & 7.1$\cdot 10^{-5}{}^{\ +16\%}_{\ -13\%}$ & 1.2$\cdot 10^{-3}{}^{\ +26\%}_{\ -19\%}$ & 2.8$\cdot 10^{-4}{}^{\ +24\%}_{\ -18\%}$ & 1.1$\cdot 10^{-3}{}^{\ +27\%}_{\ -20\%}$ & 1.1$\cdot 10^{-3}{}^{\ +36\%}_{\ -25\%}$ & 7.7$\cdot 10^{-5}{}^{\ +18\%}_{\ -14\%}$ \\
       SR2 & 1.6$\cdot 10^{-3}{}^{\ +2\%}_{\ -17\%}$ & 0 & 1.0$\cdot 10^{-4}{}^{\ +25\%}_{\ -19\%}$ & 3.5$\cdot 10^{-5}{}^{\ +28\%}_{\ -20\%}$ & 1.5$\cdot 10^{-4}{}^{\ +27\%}_{\ -20\%}$ & 2.8$\cdot 10^{-4}{}^{\ +35\%}_{\ -25\%}$ & 7.2$\cdot 10^{-6}{}^{\ +18\%}_{\ -15\%}$ \\
       SSL & 0 & 1.1$\cdot 10^{-5}{}^{\ +17\%}_{\ -13\%}$ & 2.1$\cdot 10^{-4}{}^{\ +25\%}_{\ -19\%}$ & 0 & 0 & 1.7$\cdot 10^{-4}{}^{\ +36\%}_{\ -25\%}$ & 2.1$\cdot 10^{-5}{}^{\ +18\%}_{\ -14\%}$ \\
    \end{tabular}
    }
    \caption{Cross section values after the colour-octet analysis cuts for the main background contributions together with the associated scale uncertainties. These predictions assume a conservative top-tagging performance. \label{tab:xsects_after_cuts_octet}}\vspace*{.2cm}
    \resizebox{.85\textwidth}{!}{
    \begin{tabular}{c|cccccc}
       \multicolumn{7}{c}{Colour-singlet selection [fb]} \\
         & $t \bar t +$jets & $t \bar t W$ & $t \bar t Wj$ & $t \bar t Z$ & $t \bar t Zj$ & $t \bar t t \bar t$ \\ \hline
       SR1 & 35${}^{\ +20\%}_{\ -15\%}$ & 8.7$\cdot 10^{-2}{}^{\ +18\%}_{\ -14\%}$ & 3.2$\cdot 10^{-1}{}^{\ +27\%}_{\ -20\%}$ & 3.5$\cdot 10^{-1}{}^{\ +19\%}_{\ -15\%}$ & 9$\cdot 10^{-1}{}^{\ +27\%}_{\ -20\%}$ & 1.2$\cdot 10^{-1}{}^{\ +37\%}_{\ -25\%}$ \\
       SR2 & 5.6$^{\ +19\%}_{\ -15\%}$ & 1.8$\cdot 10^{-2}{}^{\ +18\%}_{\ -14\%}$ & 4.3$\cdot 10^{-2}{}^{\ +27\%}_{\ -20\%}$ & 9.8$\cdot 10^{-2}{}^{\ +19\%}_{\ -15\%}$ & 2.0$\cdot 10^{-1}{}^{\ +27\%}_{\ -20\%}$ & 6$\cdot 10^{-2}{}^{\ +37\%}_{\ -25\%}$ \\
       SSL & 0 & 0 & 1.4$\cdot 10^{-3}{}^{\ +27\%}_{\ -20\%}$ & 2.6$\cdot 10^{-4}{}^{\ +24\%}_{\ -18\%}$ & 8$\cdot 10^{-4}{}^{\ +27\%}_{\ -20\%}$ & 1.1$\cdot 10^{-3}{}^{\ -37\%}_{\ -25\%}$ \\
    \end{tabular}
    }
    \caption{Same as in table \ref{tab:xsects_after_cuts_octet} but for the colour-singlet analysis.}
    \label{tab:xsects_after_cuts_singlet}
\end{table}

For the dominant $t \bar{t}\ + $ jets background, we generate $t \bar{t}$ events in association with one, two or three jets and use the MLM~matching and merging procedure~\cite{Mangano:2006rw, Alwall:2008qv} to combine the resulting event samples. This is especially important given our reliance on large-radius AK10 jets and the necessity for such jets to have a high transverse momentum to pass the selection. Typically, we indeed require at least three parton-level objects to have $p_T > 300$ GeV, which leads to a partonic centre-of-mass energy above 1~TeV. Our analyses however also target the associated production of a new physics resonance with a top-antitop pair as well as channels involving leptonic top decays, where soft jets may be present. Without merging, the large scale separation between the high partonic centre-of-mass energy and the low $p_T$ threshold of 20~GeV for the subleading jets could thus lead to large logarithms and an apparent breakdown of perturbativity. In addition, such resulting high-energy events favour hard initial-state radiation, which increases the possibility that a light jet or radiation product mimics a boosted top and gets mistagged. To avoid double counting and ensure proper treatment of QCD emissions, we thus tune the simulation accordingly. In particular, we combine matrix elements with up to two and three additional jets in the singlet and octet cases respectively, while fixing the \lstinline{xqcut} parameter of \amc\ to 80~GeV and the \lstinline{Qcut} parameter of \py\ to 120~GeV to smoothly regulate the separation of the matrix-element and shower regimes while preserving the high-$p_T$ behaviour of the leading jets. We have checked explicitly that the final (fiducial) cross section is stable under moderate variations of these parameters. All corresponding cards are available in our Zenodo repository~\cite{darme_2025_15783920}.

Next, we generate a $t \bar{t} b \bar{b}$ background sample as associated events could pass the colour-singlet selection where we require two top-tagged jets plus two additional isolated $b$-jets. We found that this process contributes only modestly, at most 10\% (20\%) of the $t \bar{t}\ +$ jets background in the SR1 (SR2) signal region. Consequently, matching and merging are not applied, as for any other subleading background contributions, since this would yield a significantly unimportant impact while entailing a high computational cost. Lastly, we note that the $t \bar{t} W$ and $t \bar{t} W j$ background contributions are dominant for neither the SR1 nor the SR2 regions. However, we include the $t \bar{t} W j$ contribution in our analysis as the extra final-state jet may lead to a small number of events passing the SSL selection. In this case, we do not apply any matching and merging procedure again, so the overall background yield in the SSL analysis is likely conservatively overestimated.

To conclude this section, we present in Tables~\ref{tab:xsects_after_cuts_octet} and \ref{tab:xsects_after_cuts_singlet} the fiducial cross sections of our background samples after the full reconstruction and selection procedure in the different signal regions with a conservative top-tagging performance. While the $t\bar t +$ jets contribution remains dominant, it is suppressed by more than seven orders of magnitude in the octet analysis and by about four orders of magnitude in the singlet case.

\section{Projections at current and future LHC runs} \label{sec:results}
\subsection{Coloured-octet resonances}
\begin{figure}
    \centering
    \includegraphics[width=\linewidth]{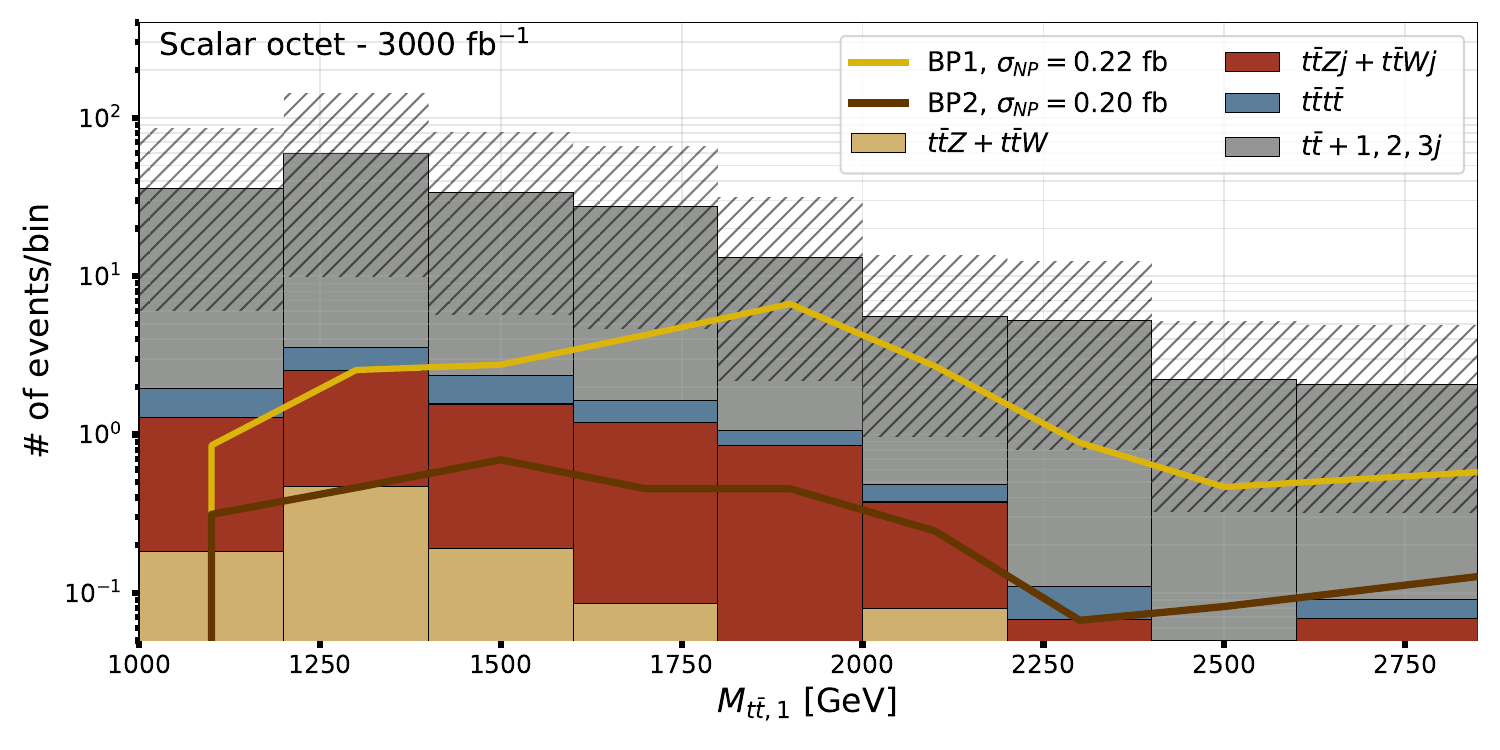}
    \caption{Signal and selected background distributions of the largest reconstructed resonance mass, for the SR1 region with conservative top-tagging and after applying our colour-octet analysis. Backgrounds are generated at LO while the signal is simulated at NLO, and the results are normalised to the HL-LHC luminosity. The hatched bands represent the cumulative scale variation uncertainties on the background and $\sigma_{\text{NP}}$ denotes the benchmark signal cross section before the selection cuts. The last bin includes the overflow. \label{fig:octet_with_bkd}}
\end{figure}

For the colour-octet case, Figure~\ref{fig:octet_with_bkd} shows the projected distributions of the largest reconstructed resonance mass $M_{t\bar{t},1}$ at the HL-LHC, in the SR1 signal region and assuming conservative top-tagging performance. The histogram includes signal predictions (simulated at NLO) for the two colour-octet benchmark scenarios BP1 and BP2, as well as the dominant background contributions (generated at LO) along with the cumulative scale variation uncertainties. The latter have been obtained by adding the different contributions linearly across the different background components, though the dominant contribution arises from $t\bar{t}+\text{jets}$ events. As discussed in Section~\ref{sec:bkd}, $t\bar{t}+\text{jets}$ production remains the leading background contribution despite being suppressed by over four orders of magnitude after our selection, owing to their initially large production cross section. This underlines the crucial role of improved top mistagging rejection as enabled by constituent-based top-tagging algorithms in order to enhance the sensitivity. 

Subdominant backgrounds include contributions from SM four-top, $t\bar{t}Z+\text{jets}$ and $t\bar{t}W+\text{jets}$ production. The SM four-top component is irreducible but has a small cross section, while the latter two processes have comparatively larger rates but are efficiently suppressed by our tagging and kinematic requirements. These backgrounds are primarily relevant for $M_{t\bar{t},1}$ values near 1~TeV, but their impact decreases at higher mass, particularly around 2~TeV where the HL-LHC bounds are expected to be found. The background distribution indeed peaks slightly above 1~TeV due to the hard $p_T$ requirements on the reconstructed AK10 jets which disfavour softer events. Consequently, TeV-scale signals are somewhat harder to distinguish from the background, though this is mitigated by the signal cross section being significantly larger than the background in this region (with rates at the fb level).

\begin{table}\renewcommand{\arraystretch}{1.25}\setlength{\tabcolsep}{10pt}
\centering
  \begin{tabular}{c|cccc|cccc}
     Top-tag. & \multicolumn{2}{c}{Optimistic} & \multicolumn{2}{c|}{Conservative} & \multicolumn{2}{c}{Optimistic} & \multicolumn{2}{c}{Conservative} \\ 
     $\mathcal{L}$ [fb$^{-1}$] & 500 & 3000 & 500 & 3000 & 500 & 3000 & 500 & 3000 \\[.2cm] 
    & \multicolumn{4}{c|}{\textbf{BP1 $-$ LO}} & \multicolumn{4}{c}{\textbf{BP2 $-$ LO}} \\ \hline
    SR1 & 0.55 & 0.21 & 0.65 & 0.25 & 10.29 & 4.21 & 11.98 & 4.59 \\
    SR2 & 0.52 & 0.13 & 0.59 & 0.17 & 5.93 & 1.88 & 6.56 & 2.03 \\
    SSL & 1.65 & 0.37 & 1.64 & 0.37 & 9.11 & 2.20 & 9.09 & 2.14 \\[.2cm]
    & \multicolumn{4}{c|}{\textbf{BP1 $-$ NLO}} & \multicolumn{4}{c}{\textbf{BP2 $-$ NLO}} \\ \hline
    SR1 & 0.59 & 0.21 & 0.72 & 0.27 & 8.28 & 3.01 & 8.66 & 3.49 \\
    SR2 & 0.59 & 0.15 & 0.67 & 0.19 & 4.16 & 1.32 & 4.59 & 1.52 \\
    SSL & 2.31 & 0.52 & 2.28 & 0.52 & 6.73 & 1.52 & 6.71 & 1.51 \\
  \end{tabular}
  \caption{Upper limits on the new physics cross section (in fb) for the colour-octet benchmark scenarios, derived from our analysis strategy across the three signal regions. Results are presented for both optimistic and conservative top-tagging assumptions and for integrated luminosities of 500 and 3000~fb$^{-1}$.}\label{tab:CSlimOct}
\end{table}

An important feature revealed by our analysis is that despite the considerable smearing associated with top-quark reconstruction, the signal retains a visible bump structure near the mass of the BSM resonance. This allows for a shape-based analysis directly on the $M_{t\bar{t},1}$ spectrum, without requiring a precise normalisation of the SM background. Theoretical uncertainties, particularly for the $t\bar{t}Z+\text{jets}$ and $t\bar{t}W+\text{jets}$ contributions, are therefore less critical to the overall sensitivity. Projected 95\% confidence level (C.L.) bounds on the signal are hence obtained by fitting the $M_{t\bar{t},1}$ distribution using the \pyhf\ framework~\cite{Heinrich:2021gyp}, with a correlated scale factor applied to account for theoretical uncertainties on the dominant $t\bar{t}+\text{jets}$ background. The resulting LO and NLO projected cross section limits for the two colour-octet benchmark points are listed in Table~\ref{tab:CSlimOct}, at integrated luminosities of $500~\mathrm{fb}^{-1}$ and $3000~\mathrm{fb}^{-1}$ and for the different signal regions defined in Section~\ref{sec:octetana}. Several qualitative trends emerge from these results. First, the search performs best when on-shell BSM resonances can be pair-produced (\ie\ for masses around or below 2~TeV). Second, the full reconstruction of the four-top final state yields stronger limits compared to the other explored strategies. In contrast the SSL channel, while benefiting from low background, suffer from lower signal yields and hence reduced sensitivity. Finally, the performance of the top tagger has a sizeable impact, with more optimistic mistagging assumptions improving the exclusion limits by up to 20\%.

\begin{figure}
    \centering
    \includegraphics[width=0.45\columnwidth]{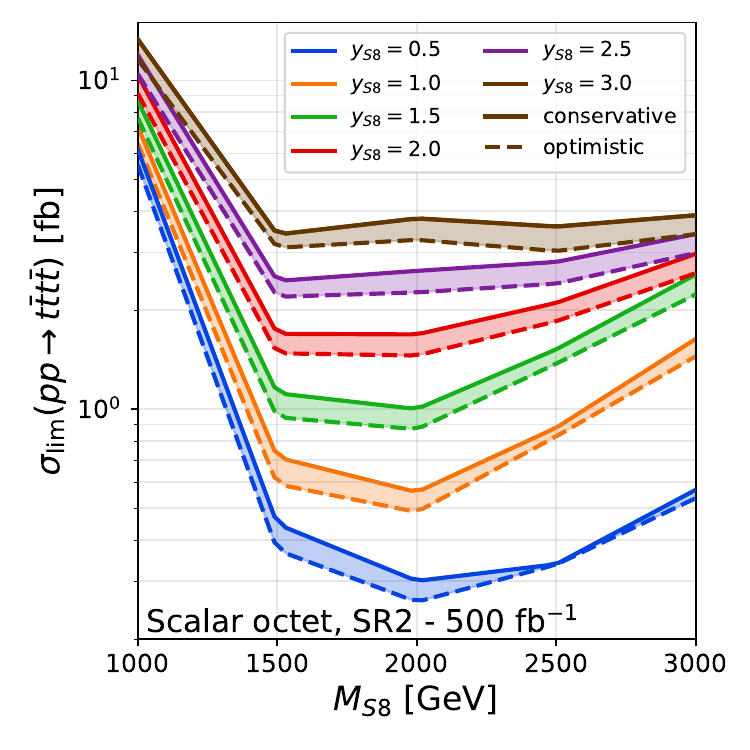}\hfill
    \raisebox{0.4cm}{\includegraphics[width=0.54\linewidth]{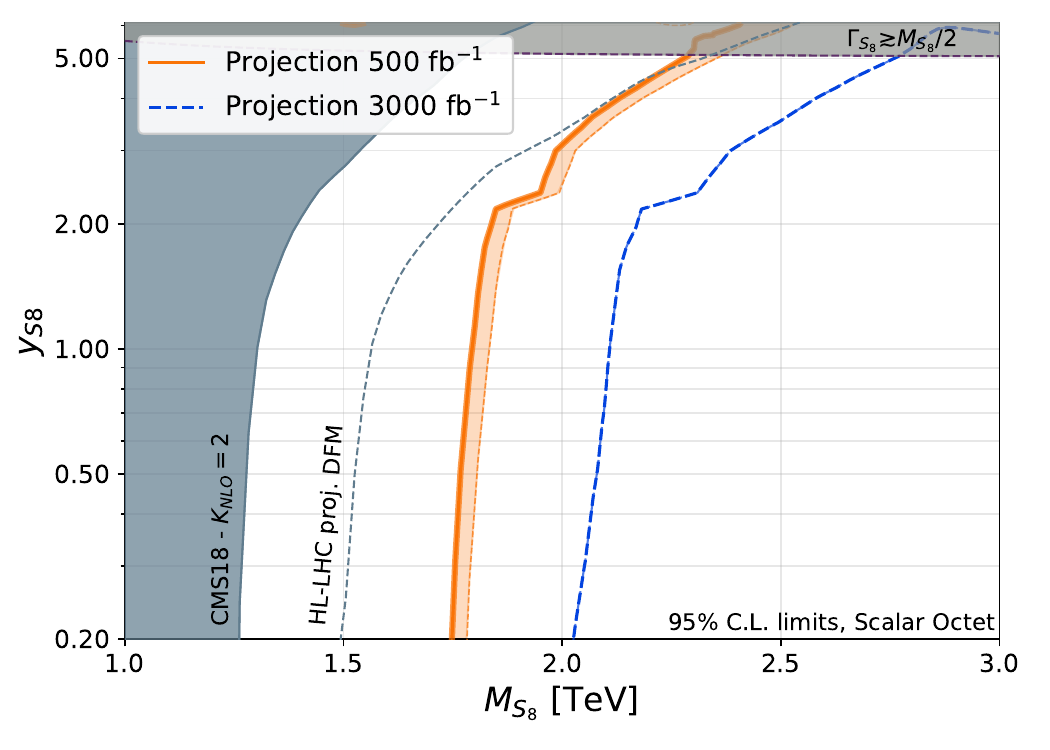}}
    \caption{\textit{Left panel} -- Expected cross section limits as a function of the colour-octet resonance mass, for several values of the coupling to top quarks and assuming a branching ratio of 1. Results are shown for the SR2 signal region at 500~fb${}^{-1}$, with the solid lines corresponding to the conservative top-tagging assumption and the dashed lines to the optimistic one. \textit{Right panel} -- Projected 95\% C.L. exclusion regions in the colour-octet mass-coupling plane for integrated luminosities of 500~fb$^{-1}$ (orange) and 3000~fb$^{-1}$ (dashed blue). Solid and dashed orange lines correspond to conservative and optimistic top-tagging assumptions, respectively, and we compare these projections to the currently excluded region (dark shading) obtained from a recast of the CMS-TOP-18-003 analysis~\cite{Darme:2021gtt, Fuks:2021zbm, DVN/OFAE1G_2020} and its naive extrapolation to 3000~fb$^{-1}$~\cite{Araz:2019otb} (dashed grey) using an approximate and overestimated $K$-factor of 2 and LO simulations for the signal. The light grey area at large couplings indicates the region where the resonance width becomes large, making our approach unreliable.}
    \label{fig:limits_scalarOctet}
\end{figure}

In the left panel of Figure~\ref{fig:limits_scalarOctet}, we present the expected cross section limits in the SR2 signal region as a function of the resonance mass for different values of the coupling to top quarks. As expected, the optimal sensitivity is achieved when pair production of the colour octet dominates, leading to a final state topology typically featuring four highly boosted top quarks. For larger values of the top-quark coupling, both the resonance width and the contribution from the associated production mode ($pp\to t\bar{t} \oct$) increase, which in turn degrades the efficiency of the search. The right panel of the same figure shows the projected 95\% C.L. exclusions in the mass-coupling plane representing the parameter space of the scalar colour-octet simplified model. Here, the NLO signal cross section is obtained by a linear interpolation as described in Section~\ref{sec:Kfact}.

Our NLO projections are in good agreement with the previous LO estimates from~\cite{Darme:2024epi}, although the earlier analysis neglected correlations between the invariant masses of the two scalar octets. In this work, we adopt a more conservative approach by considering only the largest reconstructed invariant mass per event, which partially compensates the increase in signal strength from the inclusion of NLO corrections. As a result, our analysis excludes colour-octet scalars with masses up to approximately 2~TeV, even for moderately small Yukawa couplings. As is typical for limits on BSM resonances produced via QCD-driven pair production, the projected exclusions extend down to arbitrarily small values of $y_{\oct}$ provided that the branching ratio to top quarks remains dominant and that the resonance decays promptly. Projections for more realistic scenarios where the resonance also decays into other SM particles can be readily obtained by rescaling the predicted cross section in Figure~\ref{fig:CS} and comparing it to the cross section limits in the left panel of Figure~\ref{fig:limits_scalarOctet}. Finally, we emphasise that the search strategy employed here is not fully optimised, particularly regarding the pairing method used to reconstruct the resonance mass from four-top final states. Given the substantial advances made by experimental collaborations in analogous contexts (such as double Higgs production with a $b\bar{b}b\bar{b}$ final state), we anticipate that future HL-LHC analyses could significantly improve upon our simplified approach, potentially extending the sensitivity well beyond the 2~TeV mass range.

\subsection{Colour-singlet resonance}
\begin{figure}
    \centering
    \includegraphics[width=\linewidth]{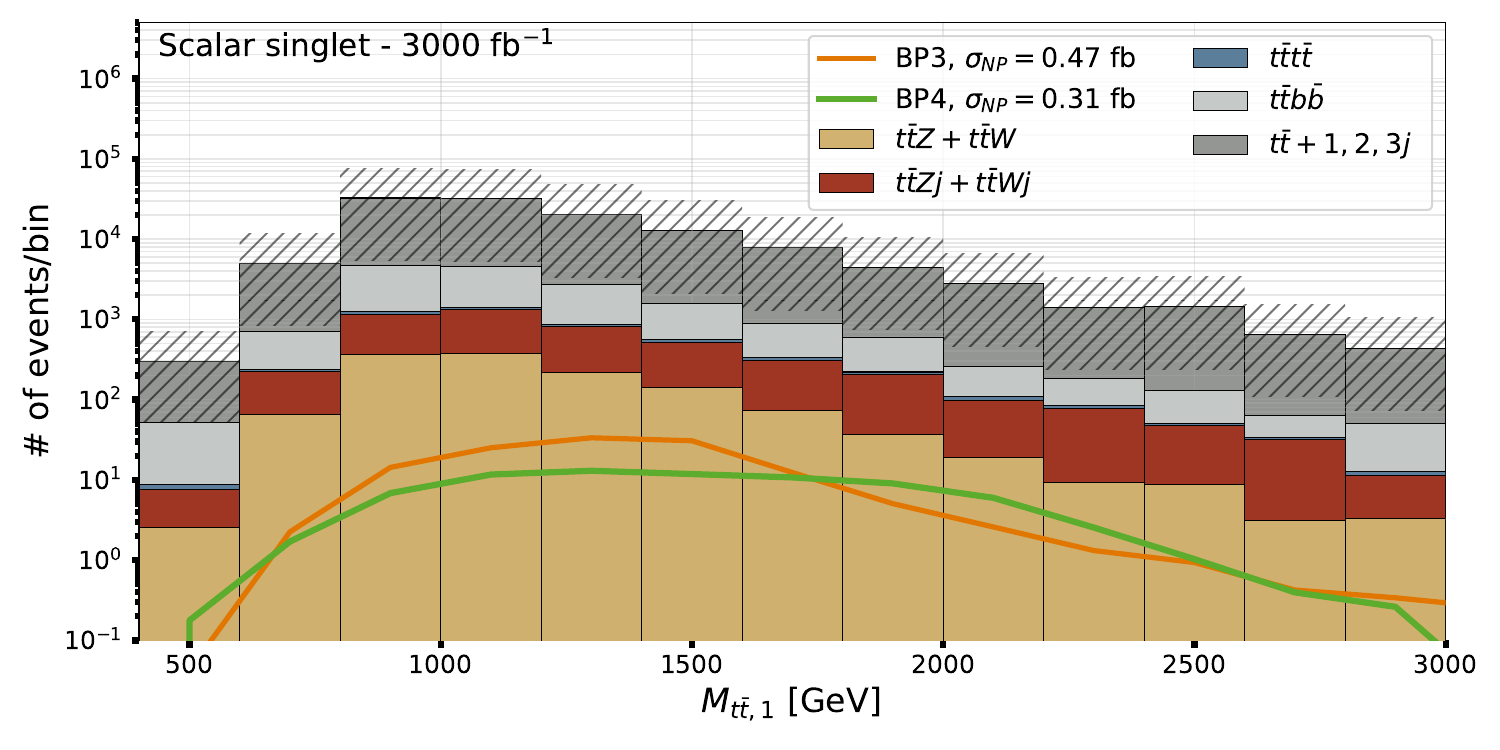}\\ \vspace{.2cm}
    \includegraphics[width=\linewidth]{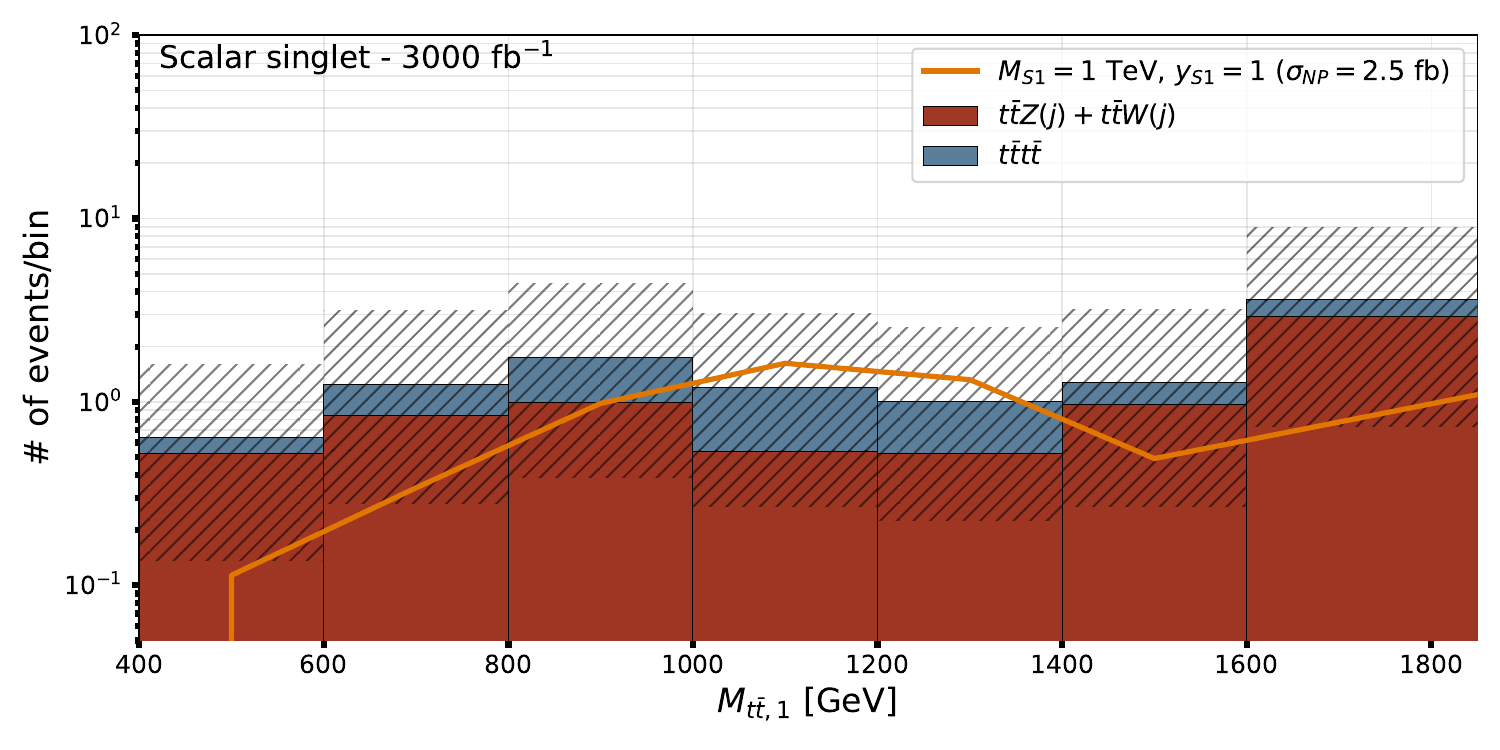}
    \caption{Signal and selected background distributions of the largest reconstructed resonance mass, for the SR1 (top) and SSL (bottom) regions with conservative top-tagging and after applying our colour-singlet analysis. Backgrounds are generated at LO while the signal is simulated at NLO, and the results are normalised to the HL-LHC luminosity. The hatched bands represent the cumulative scale variation uncertainties on the background and $\sigma_{\text{NP}}$ denotes the benchmark signal cross section before the selection cuts. The last bin includes the overflow.
    \label{fig:singlet_with_bkd}}
\end{figure}

For the colour-singlet analysis, predictions for the reconstructed invariant mass distribution of the signal at the HL-LHC are shown in Figure~\ref{fig:singlet_with_bkd} for the SR1 (top row) and SSL (bottom row) signal regions, using conservative top-tagging. The main backgrounds are also displayed, along with their cumulative scale variation uncertainties. In the SR1 region, the dominant background remains stemming from $t\bar{t}+\text{jets}$ production, like in the colour-octet case. However, the overall background level is significantly higher since the looser top-tagging requirements motivated by the typically lower $p_T$ of top quarks in the signal allow more background events to pass the selection. As in the octet analysis, our strategy nevertheless suppresses low-$p_T$ background events, which results in a peak around 1~TeV in the $M_{t\bar{t},1}$ invariant mass distribution. We emphasise that due to the specific selection cuts used in the singlet analysis, we additionally include the $t\bar{t}b\bar{b}$ background although it contributes at only the $\sim 10\%$ level compared to the dominant $t\bar{t}+\text{jets}$ background. As expected, the signal resonance is more strongly smeared in this analysis compared to the octet case, reflecting the difficulty of selecting the correct top quarks originating from the BSM decay. This suggests that more advanced reconstruction strategies may be necessary to identify a scalar top-philic singlet resonance efficiently. As a first step in this direction, we present in Appendix~\ref{appendixDiffDist} a series of differential distributions for relevant kinematic variables.

\begin{table}\renewcommand{\arraystretch}{1.25}\setlength{\tabcolsep}{10pt}
\centering
  \begin{tabular}{c|cccc|cccc}
     Top-tag. & \multicolumn{2}{c}{Optimistic} & \multicolumn{2}{c|}{Conservative} & \multicolumn{2}{c}{Optimistic} & \multicolumn{2}{c}{Conservative} \\
     $\mathcal{L}$ [fb$^{-1}$] & 500 & 3000 & 500 & 3000 & 500 & 3000 & 500 & 3000 \\[.2cm]
     & \multicolumn{4}{c|}{\textbf{BP3 $-$ LO}} & \multicolumn{4}{c}{\textbf{BP4 $-$ LO}} \\ \hline
    SR1 & 9.53 & 3.89 & 9.61 & 3.92 & 6.83 & 2.77 & 6.97 & 2.83 \\
    SR2 & 5.45 & 2.20 & 5.89 & 2.39 & 4.10 & 1.67 & 4.33 & 1.76 \\[.2cm]
     & \multicolumn{4}{c|}{\textbf{BP3 $-$ NLO}} & \multicolumn{4}{c}{\textbf{BP4 $-$ NLO}} \\ \hline
    SR1 & 10.43 & 4.26 & 10.56 & 4.31 & 8.78 & 3.57 & 8.83 & 3.60 \\
    SR2 & 6.60 & 2.57 & 8.17 & 2.91 & 4.62 & 1.87 & 5.30 & 2.16 \\
  \end{tabular}
  \caption{Same as in Table~\ref{tab:CSlimOct} but for the colour-singlet benchmark scenarios.} \label{tab:CSlimSing}
\end{table}

The projected 95\% C.L. cross section limits at $500~\text{fb}^{-1}$ and $3000~\text{fb}^{-1}$ are reported in Table~\ref{tab:CSlimSing} for the two BP3 and BP4 signal scenarios introduced in Section~\ref{sec:lags}. Like for the octet analysis, the best sensitivities for resonance masses above 1.5~TeV are obtained in SR2 where more top quarks are fully reconstructed compared to SR1. Furthermore, we do not report the obtained limits from the SSL region in the table due to the large associated Monte Carlo statistical uncertainties: generating sufficiently large samples proved computationally intensive given the low signal acceptance. However, it is not necessary in light of the slightly stronger limits obtained for the SR2 signal region. We also verify that the differences between LO and NLO signal shapes are negligible in terms of their impact on the projected limits. 

\begin{figure}
    \centering
    \includegraphics[width=0.45\columnwidth]{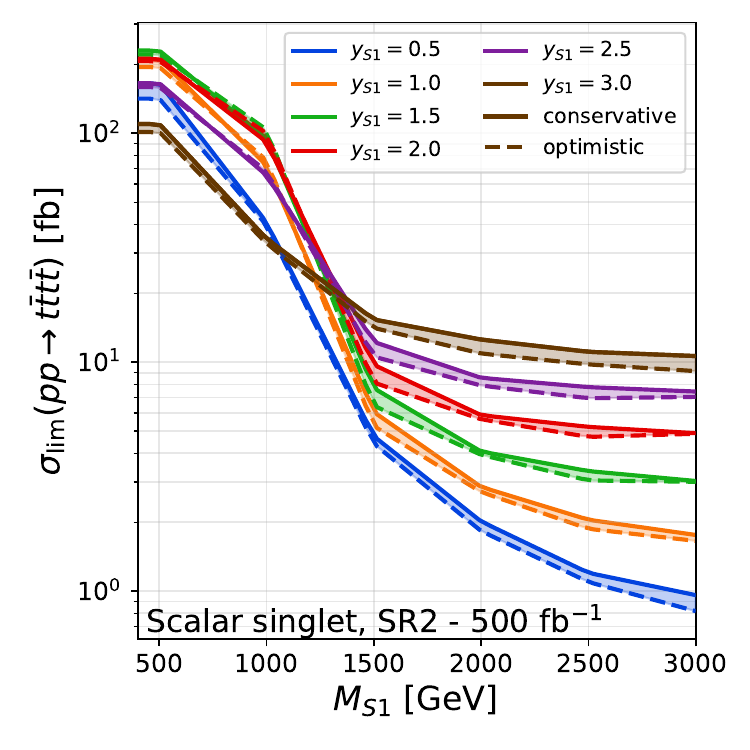}\hfill
    \raisebox{0.8cm}{\includegraphics[width=0.54\linewidth]{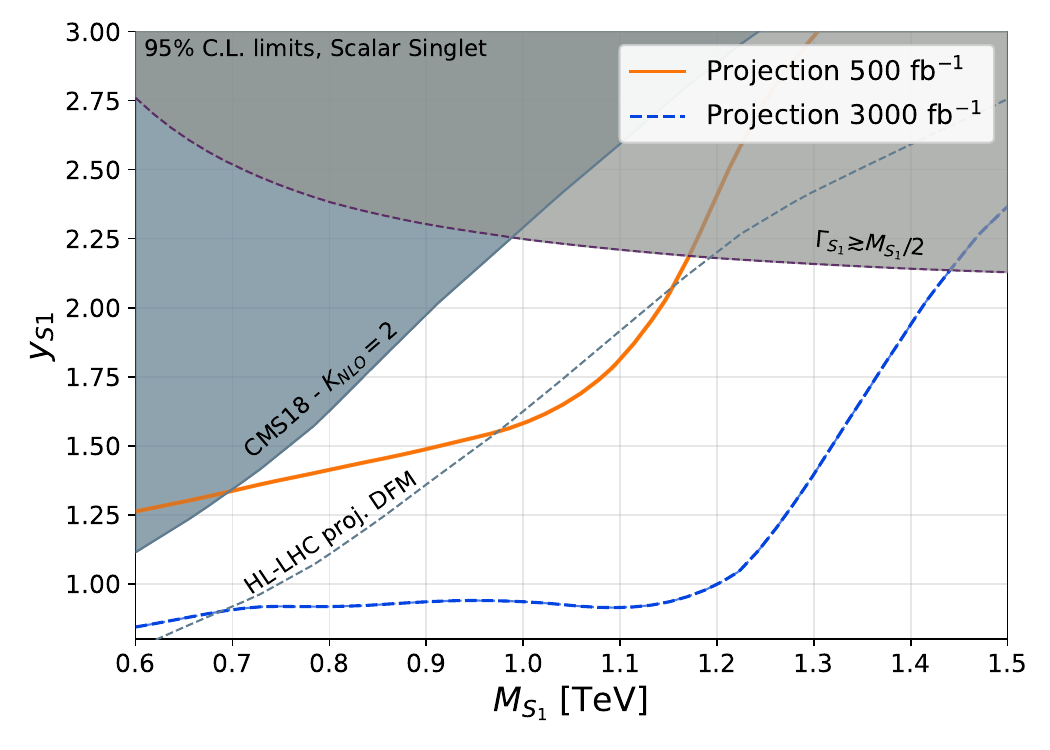}}
    \caption{\textit{Left panel} -- Expected cross section limits as a function of the colour-singlet resonance mass, for several values of the coupling to top quarks and assuming a branching ratio of 1. Results are shown for the SR2 signal region at 500~fb${}^{-1}$, with the solid lines corresponding to the conservative top-tagging assumption and the dashed lines to the optimistic one. \textit{Right panel} -- Projected 95\% C.L. exclusion regions in the colour-singlet mass-coupling plane for integrated luminosities of 500~fb$^{-1}$ (orange) and 3000~fb$^{-1}$ (dashed blue) and conservative top-tagging assumptions, using the SSL signal region. We compare these projections to the currently excluded region (dark shading) obtained from a recast of the CMS-TOP-18-003 analysis~\cite{Darme:2021gtt, Fuks:2021zbm, DVN/OFAE1G_2020} and its naive extrapolation to 3000~fb$^{-1}$~\cite{Araz:2019otb} (dashed grey), using a $K$-factor of 2 and LO simulations for the signal. The light grey area at large couplings indicates the region where the resonance width becomes large, making our approach unreliable.\label{fig:limits_scalarSinglet}}
\end{figure}

The projected cross section limits as a function of the singlet mass and coupling strength are shown in the left panel of Figure~\ref{fig:limits_scalarSinglet}. At large masses, the signal topology features increasingly boosted top quarks, which enhances the search efficiency and significantly improves the exclusion reach. For instance, the projected limit is stronger by nearly two orders of magnitude when the singlet mass increases from 1~TeV to 2~TeV. Conversely, larger values of the top-quark coupling lead to a broader resonance and enhance the relative contributions of off-shell production channels, which both degrade the sensitivity. Below the 1.5~TeV threshold, the analysis becomes markedly less effective. This is partly due to the reduced boost of the top quarks which makes their reconstruction more difficult, and partly because of the background driven by the $t\bar{t}+\text{jets}$ contribution which yields a peak in the $M_{t\bar{t},1}$ distribution around 1~TeV, thus mimicking the expected signal shape. In this region, we therefore choose to revert to the SSL search strategy which benefits from nearly background-free conditions. The lower panel of Figure~\ref{fig:singlet_with_bkd} shows the predicted invariant mass distribution for the signal in this channel, that we thus use for limit setting.

The projected 95\% C.L. exclusions for the scalar colour-singlet simplified model are displayed in the right panel of Figure~\ref{fig:limits_scalarSinglet}. As described in Section~\ref{sec:Kfact}, the limits are obtained by interpolating the LO cross section logarithmically and multiply it by a global $K$-factor interpolated linearly from an NLO-to-LO ratio grid. As above mentioned, we use the SSL analysis as it provides the best sensitivity in the relevant region of the parameter space, where perturbative values of $y_{\sing}$ only allow resonance masses up to about 1.5~TeV to be probed at the HL-LHC. Notably, even with a luminosity of $500~\text{fb}^{-1}$, our analysis improves upon existing searches for resonance masses near 1~TeV. At higher luminosities, the reach increases further, potentially allowing BSM particles with top-quark couplings comparable to the SM Higgs to be excluded up to $\sim 1.2$~TeV.

\begin{table}\renewcommand{\arraystretch}{1.35}\setlength{\tabcolsep}{10pt}
  \centering
  \begin{tabular}{c|cccc|cccc}
    & \multicolumn{4}{c|}{Singlet analysis} & \multicolumn{4}{c}{Octet analysis} \\
     Top-tag. & \multicolumn{2}{c}{Optimistic} & \multicolumn{2}{c|}{Conservative} & \multicolumn{2}{c}{Optimistic} & \multicolumn{2}{c}{Conservative} \\ 
     $\mathcal{L}$ [fb$^{-1}$] & 500 & 3000 & 500 & 3000 & 500 & 3000 & 500 & 3000 \\ \hline
    SR1 & 53.69 & 22.17 & 52.13 & 21.31 & 51.65 & 23.76 & 63.79 & 29.26 \\
    SR2 & 76.17 & 28.69 & 72.27 & 29.33 & 34.38 & 10.37 & 34.70 & 12.08 \\
    SSL &  9.8 & 3.3 & 10.1 & 3.2 & 15.6 &  3.7 & 15.8 & 3.7 \\
  \end{tabular}
  \caption{Upper limits on the colour-singlet production cross section (in fb) for a scenario with $M_{\sing}=1$ TeV and $\Gamma_{\sing}/M_{\sing}= 0.1 $, derived using both analysis strategies discussed in this work. Results are presented for optimistic and conservative top-tagging performances at integrated luminosities of 500 and 3000 fb$^{-1}$. \label{tab:CSlimSingCompare}}
\end{table}

The SR1 and SR2 conservative singlet analyses discussed earlier assumes that top-tagging is applied only to the most boosted AK10 jets. In principle, if full top-tagging could be achieved, a significant improvement of the analysis strategies would be expected,  even for resonance masses of about 1~TeV. To illustrate this point, we perform a comparison using a more optimistic reconstruction strategy similar to the one used in the octet case, and apply it to the signal originating from a scalar singlet scenario with $M_{\sing} = 1$~TeV and $y_{\sing} = 1$ (or equivalently $\Gamma_{\sing}/M_{\sing}=0.1$). The corresponding results are shown in Table~\ref{tab:CSlimSingCompare}. In this scenario, the SR2 limit improves by a factor of more than two, while the SR1 and SSL ones show no significant change. Despite this, the SSL region remains the most promising strategy across all benchmarks, further justifying its use in our final limit projections. We remind that the SSL results in Table~\ref{tab:CSlimSingCompare} carry a statistical uncertainty of approximately 20\% owing to the limited number of Monte Carlo events passing all selection cuts.

\section{Conclusion}\label{sec:conclu}

In this work, we have presented a detailed study of the prospects for discovering new top-philic BSM resonances through boosted four-top final states at the LHC and the HL-LHC. We have performed for the first time a complete NLO QCD computation of BSM-induced $t\bar{t}t\bar{t}$ production, including both on-shell and off-shell contributions. To enable these predictions, we implemented several modifications to a conventional automated software pipeline based on \fr\ and \amc\ aiming to achieve simulations that match NLO matrix elements with parton showers. We hence improved its ability to handle coloured BSM particles through a correct treatment of the renormalisation procedure, including the derivation of the necessary UV counterterms, and the introduction of a small but essential adjustment to the loop-diagram generation algorithm of \amc. The NLO corrections were found to increase the leading-order cross sections significantly, by more than 70\% in some cases, underlining the importance of their inclusion in future experimental studies.

We expanded on our previous letter~\cite{Darme:2024epi} by detailing our simulation and analysis framework relying on the use of a validated ATLAS-like detector simulation built upon standard recasting tools. Our approach focuses on directly reconstructing the invariant mass of new BSM resonances using constituent-based top-tagging instead of relying solely on traditional cut-and-count strategies in multilepton final states. Moreover, a careful treatment of the dominant backgrounds, especially the $t\bar{t}+\text{jets}$ contributions, was carried out. It includes in particular the proper matching and merging of matrix elements featuring additional jet activity with parton showers, allowing us to capture the dynamics of the high-energy tails of distributions relevant for the boosted regime. Compared to our previous work, this resulted in a more reliable estimate of background levels and improved search sensitivity.

Our main result is that at NLO, boosted four-top final states originating from pair-produced coloured BSM scalars can be robustly distinguished from background by means of bump-hunting strategies and leveraging the significant progress achieved in top-tagging techniques. This opens the possibility of probing top-philic resonances with reach comparable to that of other coloured states such as vector-like quarks or gluinos. The analysis strategies proposed here rely on simplified reconstruction algorithms, and we expect significant gains from the use of modern machine learning tools already employed in current four-top searches. For scalar singlet resonances, the limited production cross section poses a stronger challenge. Nevertheless, we have shown that complementary strategies such as using a same-sign dilepton final-state topology can remain effective for masses up to about 1.2~TeV, especially at the HL-LHC. Our comparison of top-tagging strategies further illustrates the room for optimisation in future analyses, our results hence strongly motivating the development of dedicated four-top searches for top-philic BSM resonances. 
 
\section*{Acknowledgments}

We deeply thank C.~Degrande and O.~Mattelaer for valuable discussions about NLO simulations and their implementation in \fr, NLOCT and \amc, as well as insightful discussions on the manuscript. 
Computational resources have been provided by the Consortium des Équipements de Calcul Intensif (CÉCI), funded by the Fonds de la Recherche Scientifique de Belgique (F.R.S.-FNRS) under Grant No.~2.5020.11 and by the Walloon Region. This work has received support from the European Union’s Horizon 2020 research and innovation programme under the Marie Sklodowska-Curie grant agreement No.~101028626; the work of HL and JT has been supported by the 4.4517.08 IISN-F.N.R.S convention; MM is a Research Fellow of the F.R.S.-FNRS through the grant {\it Aspirant}; HL is also supported by the start-up funding of Sun Yat-Sen University under Grant No. 74130-12255013; BF has been supported by Grant ANR-21-CE31-0013 from the \emph{Agence Nationale de la Recherche}.

\appendix
\section{Technical details} \label{sec:app}
\subsection{Full renormalisation of the simplified models} \label{appendixrenoFULL}
We use the {\sc MoGRe} package~\cite{Frixione:2019fxg} to renormalise the Lagrangian of the simplified models considered in Eq.~\eqref{eq:lags}. {\sc MoGRe} automatically introduces the renormalisation constants associated with all fields and external parameters. It then derives those related to the internal parameters of the model, following the conventions of \fr\ by truncating their dependence on other renormalised quantities at one-loop order. The code also generates counterterms for all interactions present in the Lagrangian using the full set of introduced renormalisation constants. During this process, the user retains full control over which physical quantities are renormalised and may define custom renormalisation conditions to implement a specific scheme. This last feature was however not used in this work.

In our implementation, the quarks, the gluon and the BSM resonance are renormalised on-shell. Among these, only the top quark and the BSM resonance are massive and thus require both wave-function and mass renormalisation constants instead of only a wave-function renormalisation constant. For external parameters, we renormalise the two BSM couplings to top quarks $y_{\sing}$ and $y_{\oct}$, as well as the strong coupling constant $\alpha_s$. Once the renormalised Lagrangian is constructed, the estimation of the counterterms proceeds via dimensional regularisation using \nloct. To ensure compatibility with the complex mass scheme, some manual adjustments are required. Specifically, we define complex conjugation rules and manipulate the output prior to model export into the UFO format~\cite{Degrande:2011ua, Darme:2023jdn}. We implement first the replacement
\begin{lstlisting}
  CMSConj[X] -> Conjugate[X]
\end{lstlisting}
where \lstinline{X} denotes either the BSM resonance mass, the top quark mass or the strong coupling constant. In addition, we remove complex conjugation from parameters that are assumed real such as the couplings and the renormalisation scale. Generically denoting such a parameter by \lstinline{Y}, this means the replacement
\begin{lstlisting}
  CMSConj[Y] -> Y
\end{lstlisting}
The resulting NLO UFO models are then generated using standard \fr\ functions.

For the colour-singlet scalar simplified model, additional care is required to ensure that NLO calculations with \amc\ include scalar singlets in QCD loop diagrams. This is crucial as these scalars were considered for the computation of UV counterterms by \nloct. To enforce this, we follow the procedure described in Refs.~\cite{Borschensky:2020hot, Borschensky:2021hbo} and modify the \lstinline{is_perturbating()} function in the file \lstinline{base_objects.py}, adding specifically the snippet
\begin{lstlisting}[language=Python]
if len(int.get('orders')) > 1:
    continue
## BEGIN ADDITION
if order in int.get('orders').keys() and \
   abs(self.get('pdg_code')) in [9000001]:
    return True
## END ADDITION
\end{lstlisting}
As previously indicated, these lines explicitly instruct \amc\ to include the scalar singlet (PDG code 9000001 in our implementation) in QCD loop diagrams. We verified that loops involving at least one scalar singlet were correctly generated and that UV poles cancelled as expected. For interested readers, both the full QCD+BSM renormalisation procedure detailed in this section and the more conventional QCD-only renormalisation described in Section~\ref{appendixrenoQCD} have been implemented in a {\sc Mathematica} notebook publicly available on Zenodo~\cite{darme_2025_15783920}. 

\subsection{QCD-only renormalisation of the simplified models}
\label{appendixrenoQCD}
If one wishes to use UFO models generated with the standard QCD-only renormalisation procedure implemented in \fr\ and \nloct, the main challenges arise during the process generation step and the construction of the relevant one-loop diagrams in \amc. Ensuring a consistent automated NLO calculation that properly handles both UV and IR divergences indeed requires specific care.

\begin{figure}
    \centering
    \subfloat[]{\includegraphics[width=0.3\linewidth]{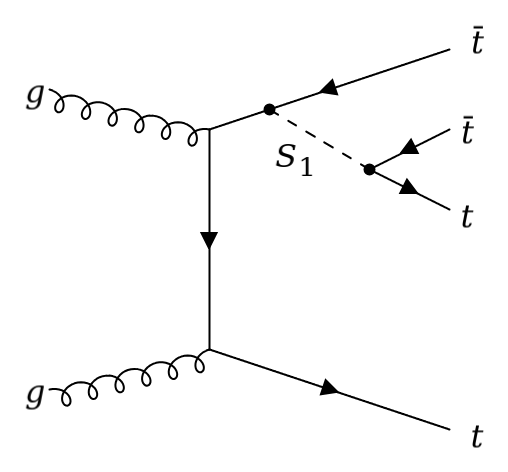}}\hfill
    \subfloat[]{\includegraphics[width=0.3\linewidth]{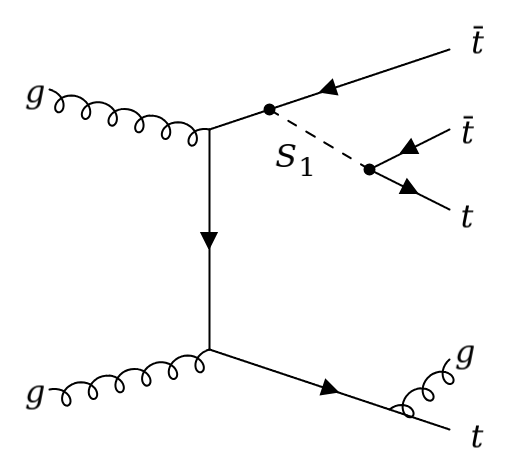}}\hfill
    \subfloat[]{\includegraphics[width=0.3\linewidth]{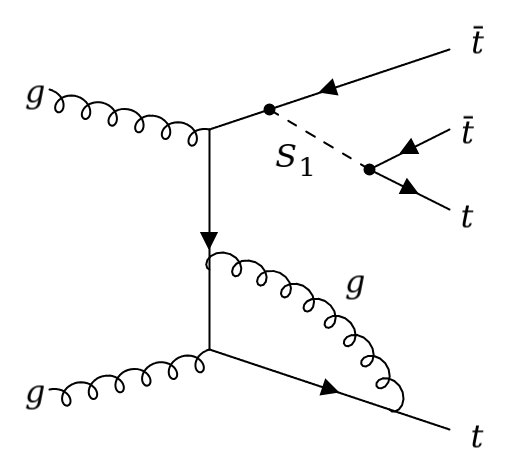}}\\[\smallskipamount]
    \subfloat[]{\includegraphics[width=0.3\linewidth]{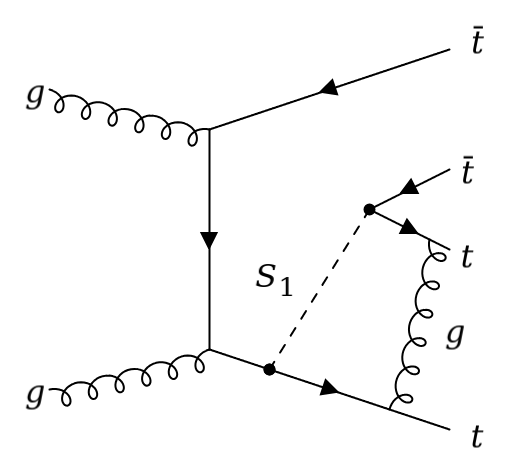}}\hfill
    \subfloat[]{\includegraphics[width=0.3\linewidth]{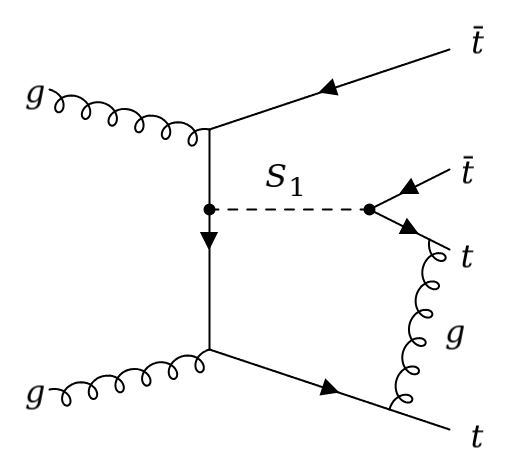}}\hfill
    \subfloat[]{\includegraphics[width=0.3\linewidth]{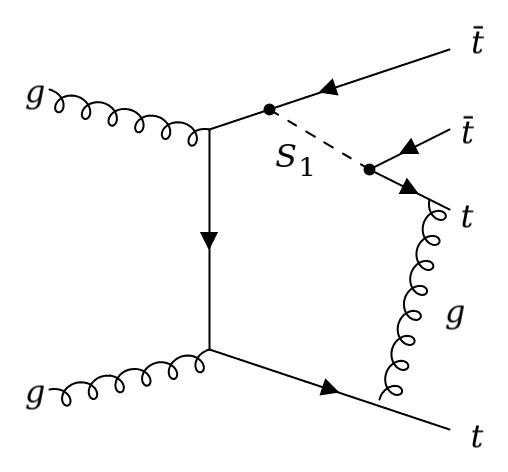}}
    \caption{Examples of Feynman diagrams contributing to the associated production process $pp\to t\bar{t}S_1$ at tree level (a) and with soft gluon radiation from a top quark (b). Relevant one-loop diagrams arise by inserting a gluon between two top quark lines. If the scalar singlet is absent from the loop, the resulting diagram is the usual QCD triangle involving only the $t\bar{t}g$ vertex (c). When a scalar propagator is instead attached to the same top quark lines as the gluon, a box diagram is obtained (d). However, the scalar propagator could also be connected to the intermediate or opposite top quark, yielding pentagonal (e) or hexagonal (f) loop.} 
    \label{fig:singletloops}
\end{figure}

In the scalar singlet model, the key subtlety concerns the treatment of IR divergences. While we could compute the process $pp \to t\bar{t}S_1$ at NLO without issue, our study focuses on the full $pp \to t\bar{t}t\bar{t}$ process where the scalar $S_1$ may appear through intermediate off-shell exchanges. In this case, the emission of a soft gluon from a top quark line in the tree-level amplitude introduces an IR divergence (see Figure~\ref{fig:singletloops}(a,b)). This divergence is cancelled by one-loop diagrams of two types. The first consists of QCD triangle diagrams with gluon exchange between top quarks (Figure~\ref{fig:singletloops}(c)), which are correctly generated along with the associated counterterms by \amc\ and \nloct. The second involves box, pentagon and hexagon diagrams with internal scalar singlet propagators (Figure~\ref{fig:singletloops}(d--f)) which are not generated by default by \amc\ since the singlet is uncharged under QCD, and which have been consistently ignored by \nloct\ (especially as they do not yield any UV divergence). However, their inclusion is necessary to cancel IR divergences. These diagrams must therefore be manually included.

\begin{figure}
    \centering
    \subfloat[]{\includegraphics[width=0.48\linewidth]{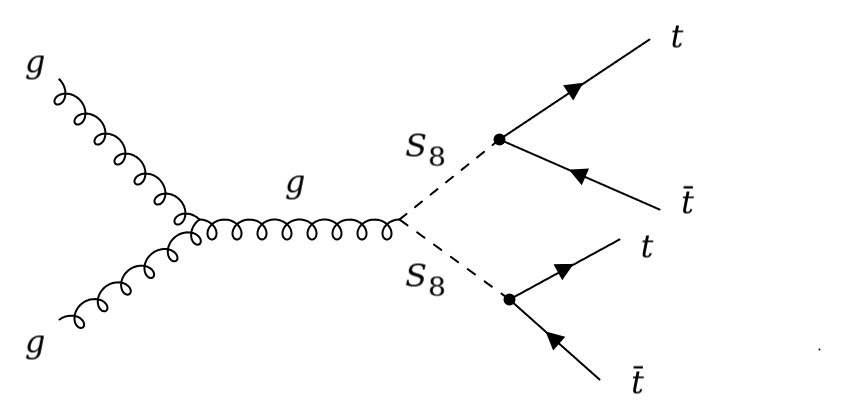}}\hfill
    \subfloat[]{\includegraphics[width=0.48\linewidth]{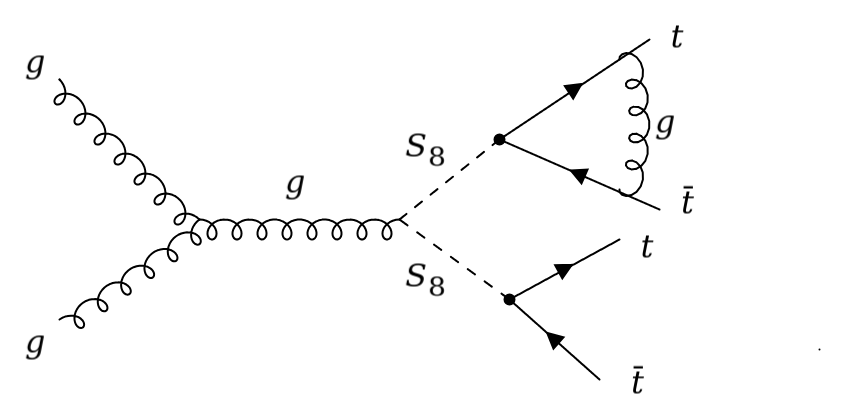}}\\[\smallskipamount]
    \subfloat[]{\includegraphics[width=0.48\linewidth]{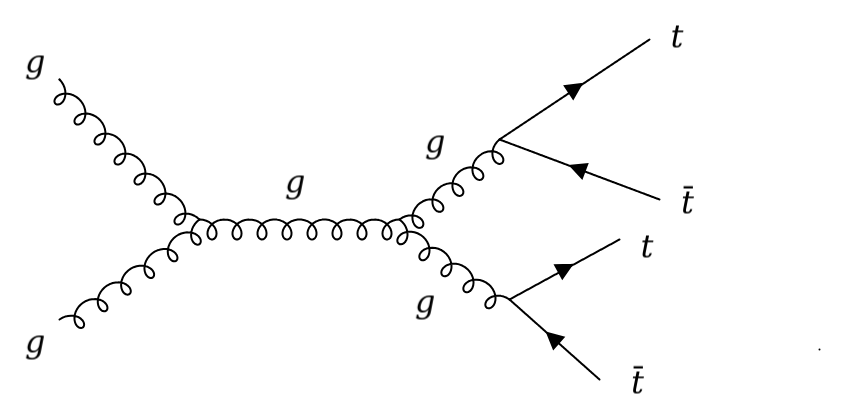}}\hfill
    \subfloat[]{\includegraphics[width=0.48\linewidth]{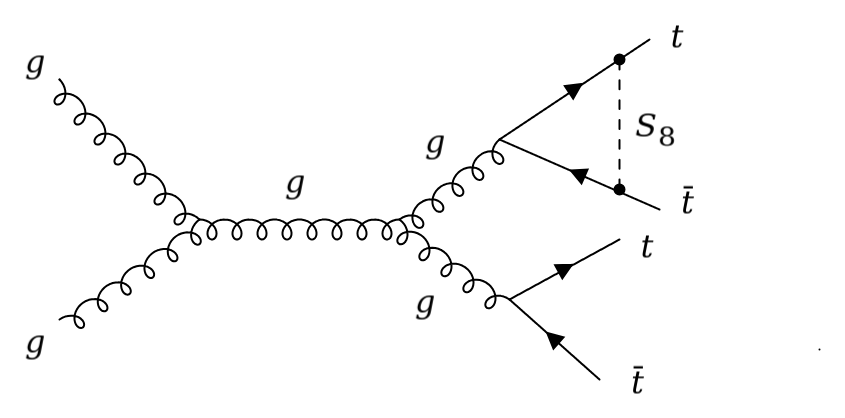}}
    \caption{Examples of Feynman diagrams for scalar octet pair production via gluon fusion at tree level (a), and with a QCD virtual gluon exchange between two top quarks (b). We additionally show a diagram for the pure SM QCD production of four top quarks (c), and with a BSM virtual exchange of a scalar octet between two top quarks (d).} 
    \label{fig:trianglettG}
\end{figure}

In contrast, the scalar octet model presents a challenge in the treatment of UV divergences. For instance, let us focus on a triangle diagram such as the one shown in Figure~\ref{fig:trianglettG}(b), which features a gluon exchange between two top quark lines and consists in a QCD correction to the tree-level diagram in Figure~\ref{fig:trianglettG}(a). At the same order, \amc\ also generates one-loop diagrams like the one in Figure~\ref{fig:trianglettG}(d), which involves a scalar octet exchange between top quarks and that corresponds to a BSM corrections to the pure QCD topology of Figure~\ref{fig:trianglettG}(c). Such a diagram contains BSM vertices that have not renormalised by \nloct, which indeed only derived the QCD counterterms. This mismatch results in uncancelled UV poles and imposes that such triangle diagrams should be removed.

We stress that the inclusion or removal of loop diagrams involving BSM resonances does not guarantee a consistent calculation. Our goal here is solely to highlight how to resolve pole cancellation issues that arise when attempting a QCD-only renormalisation of a model featuring BSM scalars. The responsibility of selecting a consistent set of diagrams lies with the user. As detailed in the main text, our main results does not rely on the QCD-only renormalisation procedure detailed in this section, but instead on the full renormalisation of both QCD and BSM sectors described in Appendix~\ref{appendixrenoFULL}.

As in the full QCD+BSM approach, a custom loop filter must be applied during loop generation in \amc. The precise implementation differs between the scalar octet and scalar singlet cases. In both cases, in the function \lstinline{user_filter()} located in the file named \lstinline{loop_diagram_generation.py}, custom filtering must be activated with:
\begin{lstlisting}[language=Python]
edit_filter_manually = True
\end{lstlisting}
For the scalar octet model, the following code should then be added to remove triangle loops involving top quarks and scalar octets but no gluons, 
\begin{lstlisting}[language=Python,showstringspaces=false]
# Apply the custom filter specified if any
if filter_func:
  try:
   valid_diag = filter_func(diag, structs, model, i)
  except Exception as e:
   raise InvalidCmd("The user-defined filter '%s' did not"%filter+
      " returned the following error:\n       > %s"%str(e))
## BEGIN ADDITION
is_incorrect_loop = (9000001 in loop_pdgs) and (6 in loop_pdgs) \
    and (21 not in loop_pdgs)
if len(diag.get_loop_lines_pdgs())==3 and is_incorrect_loop :
      valid_diag = False
## END ADDITION
\end{lstlisting}
For the scalar singlet model, the filtering logic is slightly more involved as scalar singlets must be excluded from certain loops based on their topology. This leads to the modification
\begin{lstlisting}[language=Python,showstringspaces=false]
# Apply the custom filter specified if any
if filter_func:
  try:
   valid_diag = filter_func(diag, structs, model, i)
  except Exception as e:
   raise InvalidCmd("The user-defined filter '%s' did not"%filter+
      " returned the following error:\n       > %s"%str(e))
## BEGIN ADDITION
is_loop_scalar = (9000001 in loop_pdgs)
is_loop_top = (6 in loop_pdgs)
is_not_loop_gluon = (21 not in loop_pdgs)
if len(diag.get_loop_lines_pdgs())<=3 and is_loop_scalar :
    valid_diag = False
elif len(diag.get_loop_lines_pdgs())==4 and is_loop_scalar \\
    and is_loop_top and is_not_loop_gluon :
        valid_diag = False
elif len(diag.get_loop_lines_pdgs())==5 and is_loop_scalar \\
    and is_loop_top and is_not_loop_gluon :
        valid_diag = False
elif len(diag.get_loop_lines_pdgs())==6 and is_loop_scalar \\
    and is_loop_top and is_not_loop_gluon :
        valid_diag = False
elif len(diag.get_loop_lines_pdgs())>=7 and is_loop_scalar :
    valid_diag = False
## END ADDITION
\end{lstlisting}

\section{Disentangling the pair and associated production of new scalar resonances}\label{appendixDiffDist}
In this section, we list observables that could in principle be used to distinguish between the pair production and associated production of new colour-octet top-philic resonances. These observables exploit the distinct final-state topologies that arise from the two relevant production mechanisms, and were identified from a parton-level study. However, we found that they lose most of their discriminating power after top quark reconstruction from hadron-level events, at least within our framework, due to the smearing of four-momenta and reconstructed masses. For this reason and in order to keep our analysis simple, we have not employed them (except for the first two) to obtain our main results. Instead, we always assumed that pair production was dominating in the considered colour-octet benchmark scenarios. Nevertheless, these observables may still prove useful for characterising selected signal events in real data, especially when used as inputs to a multivariate approach based on Boosted Decision Trees or Neural Networks (which can be readily integrated into our analysis framework as demonstrated in Refs.~\cite{Cornell:2021gut, Cornell:2024dki}).

The goal of these variables is thus to help determine whether selected events are compatible with the production of one or two colour-octet resonances. They probe event shape and are sensitive to the geometry and magnitude of the top quark four-momenta, attempting to assess whether one of the reconstructed top pairs originates from standard QCD interactionse.

The following list is not exhaustive but already includes a representative set of key variables.

\begin{itemize}
    \item \textbf{Difference in the top pair masses}: When two resonances are produced, the two top quark pairs in which they decay should have similar invariant masses up to reconstruction effects. This is not necessarily the case if one of the top pairs arises from QCD interactions, and we exploit this variable for the pairing of top quarks in our colour-octet analysis strategy.

    \item \textbf{Largest top transverse momentum}: In the colour-singlet analysis, we pair the two leading-$p_T$ top quarks and assume that they originate from a top-philic resonance decay. This is justified as the tops stemming from the resonance typically carry more transverse momentum than those generated through standard QCD processes.

    \begin{figure}
        \centering
        \includegraphics[width=\linewidth]{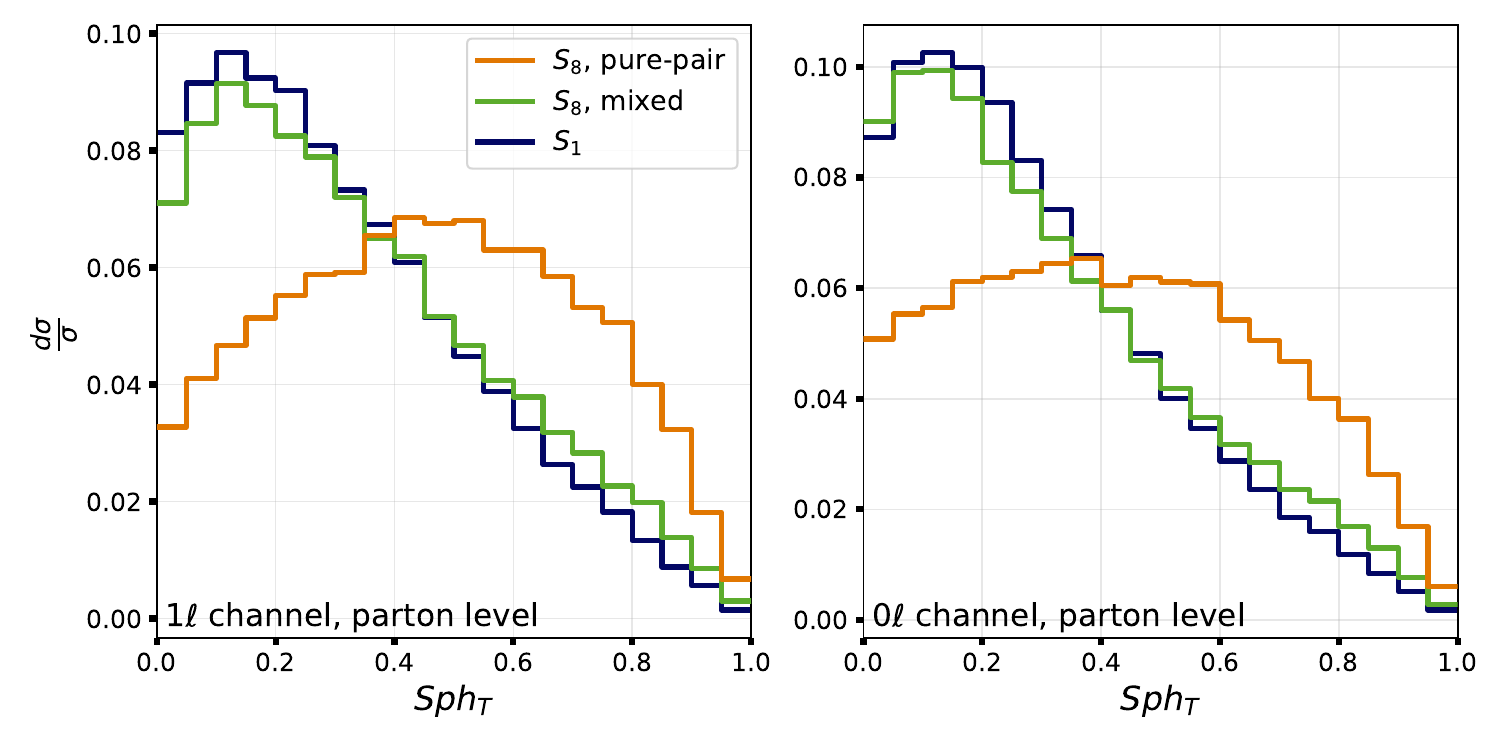}
        \caption{Normalised differential distributions of the transverse sphericity variable in the single-leptonic (left) and fully hadronic (right) channels at parton level. We compare different benchmark points for the colour-octet and colour-singlet models: the curve labeled ``$\oct$, pure-pair'' corresponds to a scenario with $M_{\oct} = 1.3$~TeV and $y_{\oct} = 0.25$ where octet pair production dominates; the ``$\oct$, mixed'' curve refers to the BP1 benchmark of Section~\ref{sec:lags} featuring significant contributions from both associated and pair production. Finally, for the singlet case, we consider the BP3 scenario of Section~\ref{sec:lags} where only associated production occurs by construction.\label{fig:sphT_lhe}}
    \end{figure}
        
    \item \textbf{Transverse sphericity}: This observable is defined as
    \begin{equation}
        \text{Sph}_T = \frac{2\lambda_2}{\lambda_1 + \lambda_2},
    \end{equation}
    where $\lambda_1 > \lambda_2$ are the eigenvalues of the transverse linearised sphericity tensor
    \begin{equation}
        M_{xy} = \frac{1}{\sum_i |\vec{p}_{T,i}|} \sum_{i=1}^{N_\text{top}} \frac{1}{|\vec{p}_{T,i}|}
        \begin{pmatrix}
            p_{x,i}^2 & p_{x,i}p_{y,i} \\
            p_{y,i}p_{x,i} & p_{y,i}^2
        \end{pmatrix}.
    \end{equation}
    Here, the momenta refer to those of the reconstructed top quarks and the variable ranges from 0 (pencil-like configurations) to 1 (isotropic distributions). Colour-octet pair production typically yields larger $\text{Sph}_T$ values, as all top quarks have four-momenta of comparable magnitude. By contrast, associated production events often feature two top quarks with significantly larger momenta than the others, resulting in a smaller $\text{Sph}_T$ value. This is illustrated in Figure~\ref{fig:sphT_lhe} for the single-leptonic (left) and fully hadronic (right) channels.

    \item \textbf{Transverse thrust}: This observable is defined as
    \begin{equation}
        \text{Thr}_T = 1 - \max_{\hat{n}_T} \frac{\sum_i |\vec{p}_{T,i} \cdot \hat{n}_T|}{\sum_i |\vec{p}_{T,i}|},
    \end{equation}
    where the sum runs over the transverse momenta of the different top quarks and $\hat{n}_T$ corresponds to the unit vector in the transverse plane that maximises the projection of all different momenta. This variable behaves similarly to transverse sphericity: values close to zero correspond to back-to-back top pairs, while values closer to $1 - 2/\pi$ signal more isotropic events.

    \item \textbf{Opening angle}: Since resonances are usually produced nearly at rest, the opening angle $\theta$ between the top quarks in which they decay tends to be close to $\pi$. This is not always the case for top quarks produced via QCD interactions. The relevant angle is defined as usual through $\cos\theta = \vec{p}_1 \cdot \vec{p}_2 / (|\vec{p}_1|\, |\vec{p}_2|)$, where $\vec{p}_{1,2}$ are the momenta of two top quarks.

    \item \textbf{Scalar triple product}: This variable is defined as
    \begin{equation}
        \frac{(\vec{p}_1 \times \vec{p}_2) \cdot \vec{p}_3}{|\vec{p}_1 \times \vec{p}_2|\, |\vec{p}_3|},
    \end{equation}
    where $\vec{p}_{1,2,3}$ are the momenta of three top quarks. This variable vanishes if the momenta are coplanar and reaches $\pm1$ when they are orthogonal. Events from associated production tend to yield values close to zero, whereas pair production favours more spread-out configurations with values closer to $\pm1$, although this behaviour also depends on the number of leptons in the final state.
\end{itemize}

\bibliographystyle{JHEP}
\bibliography{bibliography}

\end{document}